%% file: review1.tex
\documentclass[a4paper,11pt]{article}
\usepackage[]{amsmath,amscd,amssymb,ulem,floatfig}%
\usepackage[latin1]{inputenc}
\usepackage[T1]{fontenc}

\usepackage{graphicx} 

\usepackage{astron}                   

\newsavebox{\MaBoiteA}
\newsavebox{\MaBoiteB}
\newsavebox{\MaBoiteC}

\begin{document}


\begin{center}
{\huge {\bf The Bright Quasar 3C~273}}\\
\vspace{2cm}

Thierry J.-L. Courvoisier\\
INTEGRAL Science Data Centre\\
and\\
Geneva Observatory\\
\bigskip
16, Ch.\ d'Ecogia\\
CH-1290 VERSOIX\\
Switzerland
\end{center}

\begin{abstract}
  
  We review the observed properties of the bright quasar 3C~273 and
  discuss the implications of these observations for the emission
  processes and in view of gaining a more global understanding of the
  object.
  
  Continuum and line emission are discussed. The emission from the
  radio domain to gamma rays are reviewed. Emphasis is given to
  variability studies across the spectrum as a means to gain some
  understanding on the relationships between the emission components.
  
  3C~273 has a small scale jet and a large scale jet. The properties
  of these jets are described. It is also attempted to relate the
  activity in the small scale jet to that observed in the radio and
  infrared continuum.

\end{abstract}

\section{Introduction}

3C~273 is the brightest quasar on the celestial sphere. This would
suffice to qualify it to be the object of intense study over the years
since the discovery of quasars in 1963. In addition, however, 3C~273
displays most if not all the phenomena that make Active Galactic
Nuclei (AGN) such intriguing objects. Indeed, 3C~273 is a radio loud
quasar that shows large flux variations at all wavelengths and at some
epochs a non negligible polarization. 3C~273 has a small scale jet,
the features of which move away from the core at velocities apparently
larger than the speed of light and a radio structure that extends to
large distances. Whether this makes 3C~273 an archetype of AGN in
general or merely a very special case is left to the reader to assess.
What is certain, though, is that the study of 3C~273 is relevant to
all the AGN physics. The study of 3C~273 has therefore been very
actively pursued over more than 3 decades.

This study has been further made easier by the position of 3C~273 very
close to the celestial equator. This means that the object can be (and
has been extensively) observed with the most appropriate instruments
located as far North as Finland and as far South as Chile. This
privileged position on the sky has the only disadvantage that the Sun
comes close to 3C~273 on the celestial sphere during the year, hence
hampering regularly the observations.

We will review in the next pages the early work on 3C~273, analyse
then the existing data on the continuum and line emission including
their variations. We will then discuss the jet which is observed at
small and large distances from the core. The host galaxy and
environment will also be described. Some of the efforts made to
understand the physical nature of emission components and the quasar
itself will be presented.

\begin{figure}
  \includegraphics[totalheight=15cm,width=12cm]{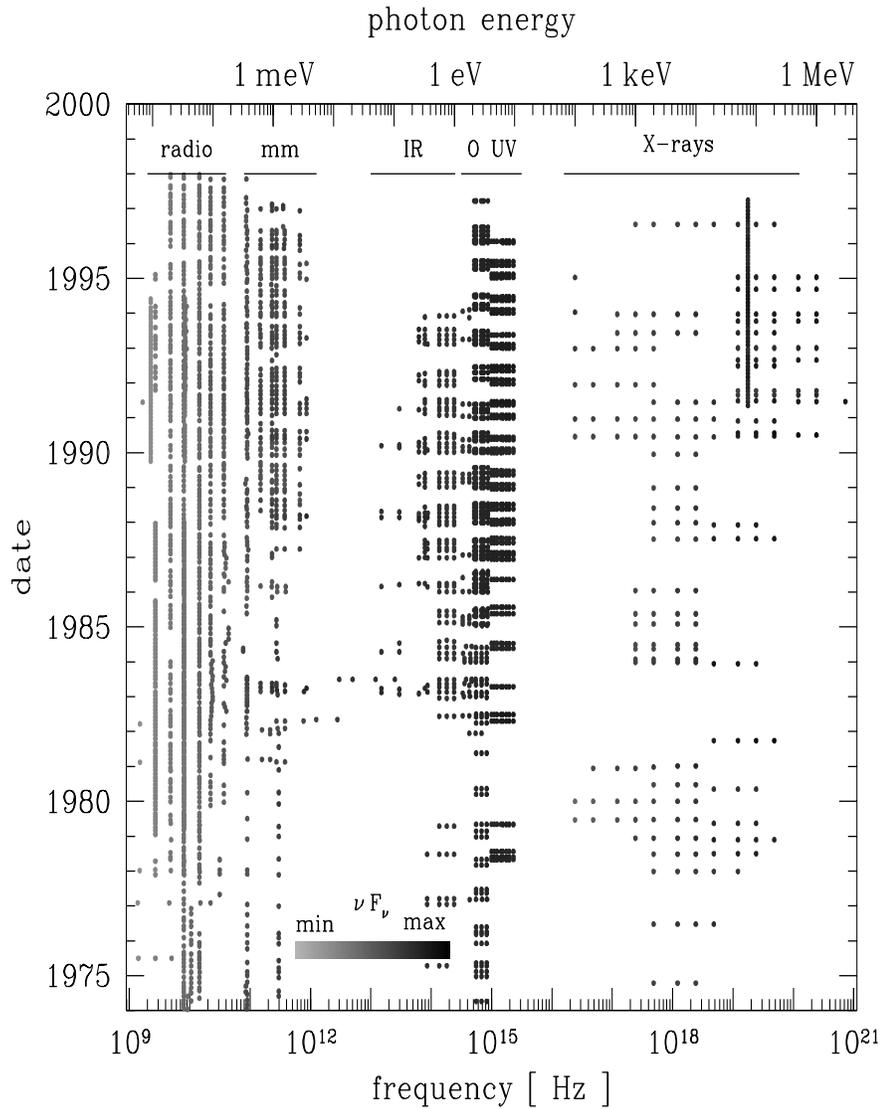}
\caption{\label{fig:theflux}The flux per logarithmic frequency
  internal $\nu \cdot f_{\nu}$ is grey coded for all the frequencies
  and all the epochs since 1975 for which data available to us have
  been obtained. (From T³rler et al.\ in preparation, see
  section~\ref{sec:continuum}). This figure shows the vast amount of
  data that have been gathered but illustrates also that most of the
  $\nu \times$ epoch plane is not covered. A good understanding of the
  object would allow us to confidently extrapolate the observed data
  to fill the gaps. This understanding is not available yet.}
\end{figure}

This review is written shortly after the switch off of the
International Ultraviolet Explorer (IUE) satellite which has been one
of the main research instruments for AGN in recent years. It is
therefore timely, as it ought to be possible to summarize the
knowledge we have including all the IUE data. Large sets of data in
other wave bands have been obtained in a coordinated way while IUE was
taking observations, as shown in Fig.~\ref{fig:theflux}. Some discussions of these
data and in particular cross correlations between the light curves in
different bands are included in the present review, although the work
is still on-going.  It will be seen that it is difficult to understand
all these data in a consistent way.

The equatorial coordinates of 3C 273 are $\alpha_{2000} =
12$\,h~29\,m~06.7\,s, $\delta_{2000} = +
02^{\circ}03^{\prime}08^{\prime\prime}$. The galactic coordinate of
the quasar are l~$= 289.95$ and b~$= +64.36$. The V magnitude
(average) is 12.9. 3C 273 is therefore a high galactic latitude
object, a clear advantage to minimize the effects of gas and dust along the line of
sight. The redshift of 3C 273 is z~$= 0.158$ \cite{veroncat95}.

\section{Discovery and early work}

3C 273 is the second radio source identified with a stellar object
\cite{Haz63} and the first for which the emission lines were
identified with a redshifted Hydrogen sequence \cite{Schmidt63}. It
was immediately clear that if the redshift is a measure of the
distance to the object its luminosity must be extraordinarily large.
\cite{Schmidt63} already recognized the presence of a faint jet-like
nebulousity. Early work on quasars has been summarized in
\cite{Burbidge67} and 2 years later by \cite{Schmidt69}.

Early spectroscopic work interpreted the emission lines already
identified in the frame of the then familiar gaseous nebulae theory
\cite{GreensteinSchmidt}. These authors found that the line emitting
gas must have a temperature of the order of 17\,000\,K and electron
densities around $10^6$cm$^{-3}$. It must be stressed that there was
then no distinction between the narrow and broad line regions. These
interpretations made it clear that the heavy element abundances must
be close to the values found in galactic nebulae, hence close to those
found in young stars. This was soon found to be a problem if the
objects were to be at the cosmological distances suggested by their
redshifts \cite{Shklovsky64}. Indeed in this case it was recognized
that the light emitted by quasars must have originated when they were
very young and it was not expected then that extensive stellar
nucleosynthesis takes place in so short a time after the formation of
structures in the expanding Universe. It now seems well established
that star formation is closely associated with the process of
accretion of matter from the galaxies to the very central regions in
which nuclear activity takes place. Thus explaining why no nuclear
activity involving only Hydrogen and Helium is observed.

Soon after the discovery of 3C 273 as an unusual object existing plate
collections were used to obtain light curves. Using the Harvard
\cite{SmithHoffleit63} and Pulkovo \cite{SharovEfremov63} plate
collections it was then realised that the object had varied by a
factor of approximately 2 in the course of the preceding decades,
although no such variations had been observed with the then modern
data. Radio flux variations were also soon found in one of the
components of 3C~273 \cite{Dent65}.

It was also noted early that the continuum spectral energy
distribution of 3C~273 (and Seyfert galaxies) might cover all the
known bands of the observable electro-magnetic spectrum. The emission
is such that roughly the same amount of energy is observed in the
different parts of the spectrum.  This remark followed the possible
detection of X-rays from 3C~273 reported by \cite{FriedmanByram67} and
the observation of an important infrared flux \cite{PachoWeymann68}.

The main observational pieces of the problems posed by quasars
(extreme luminosity, variability and hence compactness and emission
covering most of the observable electromagnetic spectrum) were thus in
place in a very short time after the first identification of point
like radio sources with red shifted optical sources.

The very unusual properties of 3C~273 and a few other quasi stellar
sources as compared to any of the astrophysical objects then known
(stars, nebulae, supernovae etc) provoked a wide variety of possible
explanations summarized in \cite{BurbidgeBurbidge67}. The necessity to
use concepts very different from those related to nuclear processes to
explain the energy requirements of Quasi Stellar Objects (QSOs), if
these are at the distances implied by their redshifts, even led to
alternative ideas to explain the redshifts. These interpretations
became less and less plausible when the association of AGN and quasars
with galaxies in which the emission and absorption line redshifts
coincide and when absorption lines due to intervening matter also at
cosmological distance were discovered. It became more and more evident
then that the energy released by matter when it is accreted into the
very deep potential well associated with supermassive black holes is
at the origin of the QSO phenomena. This idea had first been suggested
by \cite{Zel'dovichNovikov64} and \cite{Salpeter64} and has since then
become the paradigm in which the theory of Active Galactic Nuclei
(AGN) develops.

\section{Continuum Emission \label{sec:continuum}}

\begin{figure}
  \includegraphics[totalheight=15cm,width=12cm]{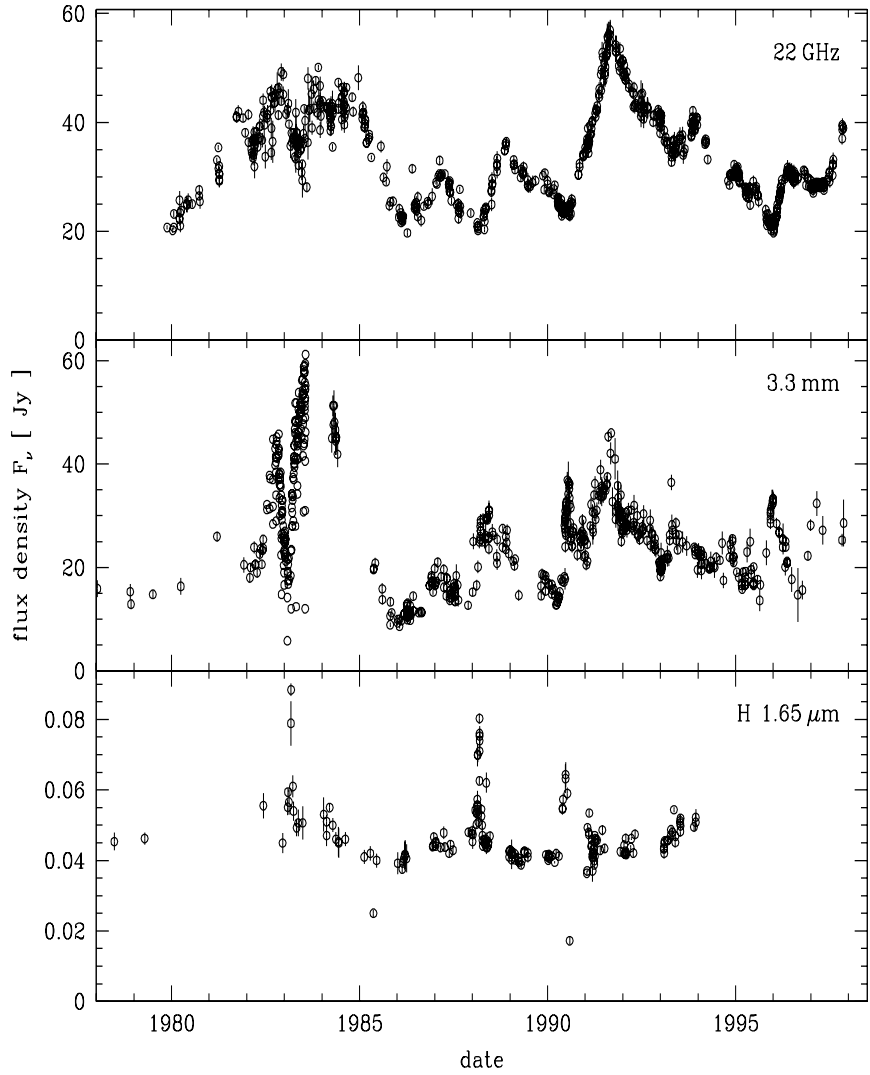}

\caption{\label{fig:lightcurves}Light curves (flux as a function of time) for a representative set of frequencies in the radio to infrared bands. These curves are (up to a factor $\nu$) vertical cuts through Fig.~\ref{fig:theflux}.}
\end{figure}

\begin{figure}
  \includegraphics[totalheight=15cm,width=12cm]{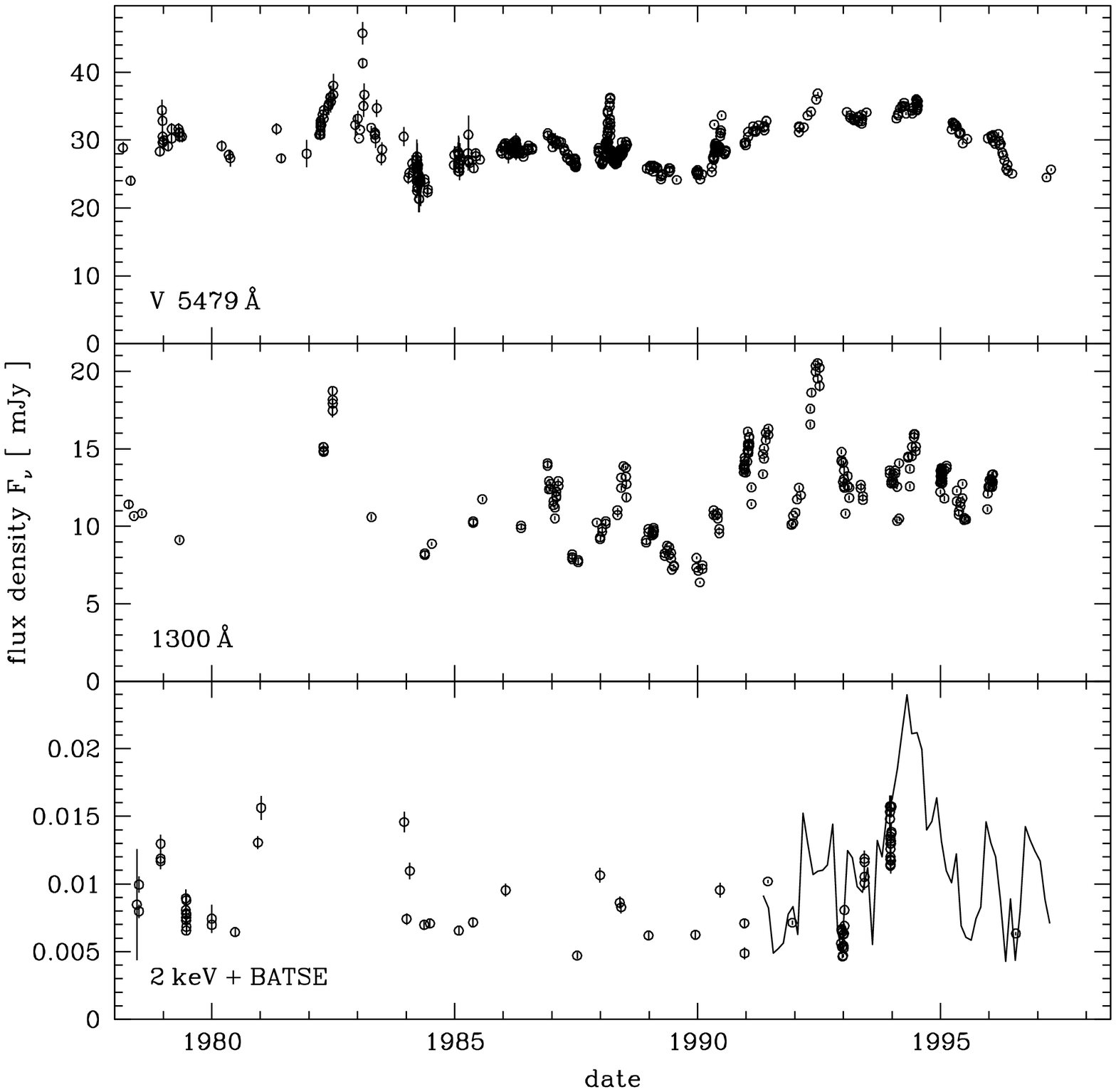}
\caption{\label{fig:Lightcurves2} Light curves in the visible ultraviolet and X-ray bands. Data from Fig.~\ref{fig:theflux}. The solid line  in the lower panel is an extrapolation to 2\,keV of the BATSE light curve using an average spectral index.}
\end{figure}

The emission in the different bands of the electro-magnetic spectrum
is described in this section, both spectral information and
variability are discussed. Figures~\ref{fig:lightcurves}
and~\ref{fig:Lightcurves2} give a representative subset of the light
curves on which the description of variability is based.  These curves
are cuts at given frequencies of the data presented in
Fig.~\ref{fig:theflux}. We give at the end of the section the average
overall spectral distribution (the projection of the data in
Fig.~\ref{fig:theflux} on the frequency axis). The data shown in
Fig.~\ref{fig:theflux} were collected by several groups, they will be
described in some details in T³rler et al. (in preparation) and made
available on the web. In the following, data used and not specifically
referenced will be drawn from this collection.

\subsection{Radio Emission}

Low frequency (between 4.8 and 14.5\,GHz) radio observations of several
quasars including 3C~273 have been performed regularly at the
University of Michigan Radio Astronomy Observatory (UMRAO). These data
have been described by \cite{Alleretal85}. A further low frequency
monitoring (2.7 and 8.1\,GHz) from 1979 to 1987 is presented in
\cite{Waltmanetal91}. The main characteristics of the low frequency
emission are a flux of about 40\,Jy at 8\,GHz with 2\% linear
polarization. The light curve shows a broad minimum that lasted
approximately 2 years around 1980. At 14.5\,GHz the flux decreased by a
factor 2 at the deepest point.

The variability in this frequency domain is further analysed in
\cite{Hughesetal92} in terms of structure functions. It is possible to
measure, using structure functions, the longest time on which
variability occurs. The structure function

$$S(\tau) = < (f(t+\tau) - f(t))^2 > = \frac{1}{\rm{N}}
\sum_{i=1}^{\rm{N}} {(f(t_i+\tau) -f(t_i))}^2, $$

N being the number of observations and $t_i$ the epoch of each
observation, is given by the average of the square of the difference
of fluxes ($f$) observed at two epochs separated by $\tau$. The
structure function thus increases as a function of $\tau$ as long as
$\tau$ is less than the maximum timescale on which the source varies.
Beyond this maximum timescale the structure function is flat, and
reflects the square of the amplitude of variations. In the case of
3C~273, the low frequency emission variability is such that the
longest relevant time scale is longer than the data span available for
the analysis, i.e.\ longer than 10 years.

At higher frequencies (from 22 to 87\,GHz) a dense and regular
monitoring of a sample of active nuclei including 3C~273 is performed
at the Metsõhovi Radio Research Station. This data set is described in
\cite{Terõsrantaetal92}. Radio emission at these frequencies shows an
increase in the amplitude of the variations as compared to the lower
frequencies. At 22 and 37\,GHz, the flux varies between 20\,Jy and
60\,Jy. This amplitude is indeed larger than the factor 2 variability observed
at lower frequencies. There are no periods during which the radio
emission can be described as "quiescent" onto which "flares" would be
superimposed. Rather, variation is the rule and not the exception.

\begin{figure}
  \includegraphics[totalheight=15cm,width=12cm]{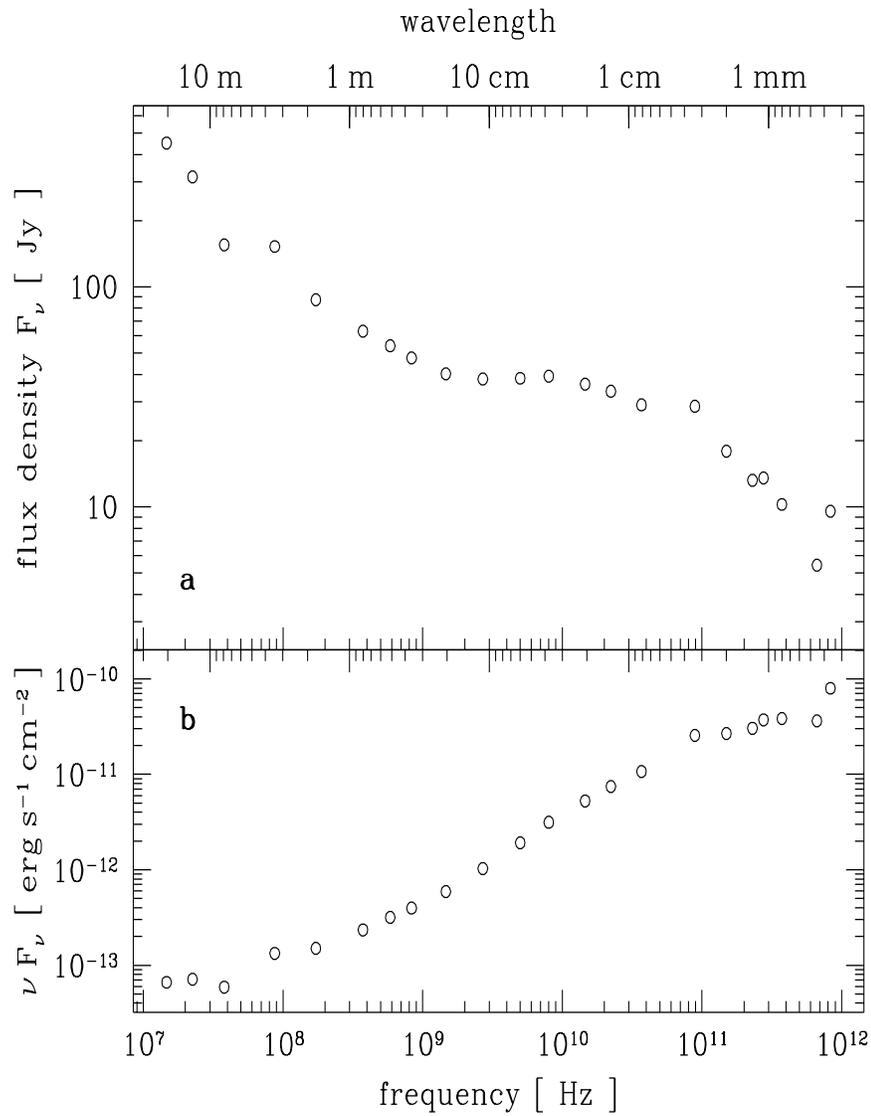}
\caption{\label{fig:average1}The average radio-millimetre
  spectrum. Data from Fig.~\ref{fig:theflux}. Top panel gives the flux
  density $f_{\nu}$ while the bottom panel gives $\nu \cdot f_{\nu}$\,, the flux per (natural) logarithmic frequency interval.}
\end{figure}

The non thermal character of the radio emission evidenced by the
spectral energy distribution (see Fig.~\ref{fig:average1}), the strong
variability observed at high radio frequencies and the polarization of
the flux are three elements that undoubtedly identify the radio
emission mechanism as synchrotron radiation. The variability of the
sources is one of the main elements to postulate that the relativistic
electrons emitting the synchrotron radiation are generated in shocks
associated with the jets (see \cite{MarscherandGear85} and references
therein). In this model shocks propagate along the jet from denser to
less dense regions. In this process, the frequency at which the
perturbed region becomes optically thin decreases with time. It
follows that the millimetre emission increases and the frequency of
the peak emission is displaced towards lower frequencies. In the next
stage, the region expands quasi adiabatically, the radiation losses
being less important. Flares are thus described by a flux increase in
the millimetre domain that propagates to lower frequencies with time.
(See T³rler et al\ in preparation for a 3-dimensional (flux versus
time and frequency) representation of the model for average flares).

The spectral energy distribution is more complex than a single power
law (Fig.~\ref{fig:average1}). In many AGN the presence of broad humps
as seen here at 10\,GHz is taken as evidence for the presence of cool
dust at large distance from the ultraviolet source of the nucleus.
Here, the large amplitude of the variability on short timescales (of
the order of a year) indicates rather the superposition of several
synchrotron components.

It is possible to describe the radio-millimetre flux variations up to
frequencies of 100\,GHz by a set of successive independent events
(T³rler et al., work in progress) in such a way that the resulting
light curves are all well described. Each event is parametrised by a
rise time and a decay, a spectrum and the time of the start of the
event and its intensity. The decomposition is performed by fitting a
set of flares (about 1 per year) simultaneously to more than 10 light
curves covering the spectrum from 0.3\,mm (1\,000\,GHz) to 10\,cm
(3\,GHz) during the last 20 years. This method is able to isolate
individual outbursts and to derive their evolution as a function of
both time and frequency. Preliminary results show that the observed
properties of a typical outburst in 3C~273 are in good qualitative
agreement with the predicted properties by shock models in
relativistic jet like those of \cite{MarscherandGear85}.

\subsection{Millimetre and infrared emission}

\begin{figure}
  \includegraphics[totalheight=15cm,width=12cm]{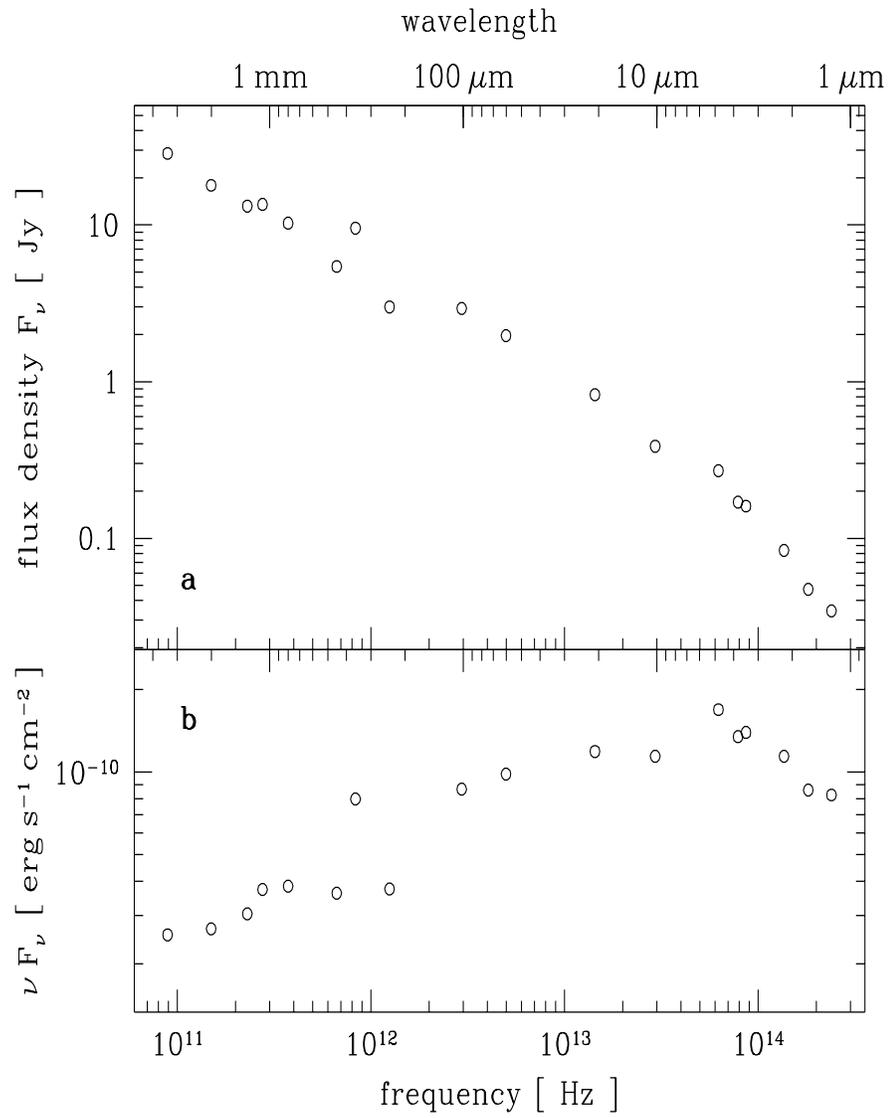} 
\caption{\label{fig:average2}The average millimetre-infrared spectrum of 3C~273. Data from 
Fig.~\ref{fig:theflux}. Panels as in Fig.~\ref{fig:average1}.}
\end{figure}

Figure~\ref{fig:average2} gives the average spectrum between $10^{11}$\,Hz and
$10^{14}$\,Hz extracted from Fig.~\ref{fig:theflux}. This continuum spectral energy
distribution is characterised by a power law of index of $0.7 \pm 0.1$
(between 10\,$\mu m$ and 100\,$\mu m$) and a "bump" around a few
microns (see below).

Early infrared data are presented in \cite{Neugebaueretal79} and in
\cite{RiekeLebofsky79}. \cite{Neugebaueretal79} note that the 3\,$\mu m$
flux of 3C~273 and other quasars is in excess of a power law that
would link the 1\,$\mu m$ and the far infrared fluxes. They suggest
that this bump (easier to see in the $\nu\cdot f_{\nu}$ representation)
around 3\,$\mu m$ may be due to the presence of heated dust, the
emission of which is superimposed on the non thermal emission that
extends smoothly from the radio domain. A similar conclusion is
reached by \cite{Allen80}. The presence of dust within the nucleus
might also explain the ratio of Ly$\alpha$ to H$\alpha$ fluxes which
is an order of magnitude less than the theoretical predictions
\cite{Hylandetal78}. Dust located within the line emitting region
could, according to these authors, lower this line ratio
through reddening effects. Indeed, even small amounts of reddening
will considerably decrease the Ly$\alpha$ flux while the H$\alpha$
flux will remain nearly unperturbed. Continuum observations suggest,
however, that this dust does not redden the continuum in the same way.
There is in fact no indication of substantial reddening in either the
UV domain or the X-ray domain (see below) in excess to that caused
within our Galaxy.

\subsubsection{Variability of the millimetre-IR emission.}

The first observation of a millimetre outburst in 3C~273 is described
in \cite{Robsonetal83} who followed the spectral energy distribution
throughout the event. This was subsequently interpreted by the model
based on a shock in an expanding jet mentioned
above~\cite{MarscherandGear85}. Although the model is probably too
simplistic to be directly applicable, it remains one of the main tools
to understand the radio and millimetre variability of 3C~273 and other
sources.

The millimetre observations of 3C~273 in 1986 showed
\cite{Robsonetal86} that the sub-millimetre flux could also decrease to levels
well below that normally observed. This happenened while the infrared
flux remained constant at wavelengths shorter than 10\,$\mu m$. The
radio-millimetre emission of 3C~273 is thought to be due to
synchrotron emission. Energetic synchrotron emitting electrons
radiating at high frequencies loose their energy faster than less
energetic electrons radiating at lower frequencies. The behaviour observed in
1986  is therefore in contradiction with expectations based
on synchrotron emission. This result firmly established the
presence of another component in the infrared continuum of 3C~273. The
small amplitude of the variations in the near infrared and the
arguments described above strongly suggest that this component is due
to dust close to the sublimation temperature.

The infrared emission of 3C~273 is due to 2 very different components.
On one side, the dust that has already been mentioned and on the other
a rapidly flaring component that is observed only during short but
violent events (see Fig.~\ref{fig:lightcurves}). This activity was
observed for the first time in 1988 \cite{Courvoisieretal88}. The flux
variations observed then are such that, assuming isotropic emission,
the luminosity changes are about $6 \cdot 10^{40}$\,ergs s$^{-2}$. The
polarization during the flare (few percent) was much larger than
during quiescent periods. Both the strong variations and the high
polarization imply that this flaring component is of synchrotron
origin. Using the variability timescale in the K band and assuming
that the emission is due to electrons cooling through synchrotron
emission, \cite{Courvoisieretal88} deduced that the magnetic field was
of the order of 1 Gauss and the Lorentz factor of the electrons
emitting the flare of the order of $10^4$. It was later established
that these flares may be at the origin of new components in the VLBI
jet of 3C~273 (see Sect.~\ref{subsec:VLBI_Jet}). A similar flare
probably occurred in 1990 (see Fig.~\ref{fig:lightcurves}), as can be seen in the long
term light curves. It is not possible to study the duty cycle or the
frequency of this activity, as the flares are very short and the flux
during the flares is extremely variable. The flares are therefore
easily missed in long term sets of observations which do not have a
sufficiently dense sampling to systematically catch the events.

It is interesting to note that the energy radiated during the
synchrotron flare of 1988 is of the order of $10^{51}$\,ergs (an
isotropic flux of 20 mJy in a band width of 1\,$\mu$ in the near
infrared at a distance of 1\,Gpc for about 1 day) and to consider
whether the radiated energy could have been stored in the magnetic
field. The energy available in a magnetic field of about one Gauss as
deduced in \cite{Courvoisieretal88} over a volume of a few light days across
is of the order of $10^{46}$\,ergs, insufficient to explain the flare.
Another possible energy source is the kinetic energy of mass ejection.
Assuming a mildly relativistic velocity of 0.1\,c decelerated in about
one day as the flare energy source one estimates that about $6 \cdot
10^{26}$\,g must be decelerated and produce synchrotron radiation with
a 100\,\% efficiency to explain the observed luminosity variations.
This is the mass accreted by the central black hole every second (see
Sect. 8). These estimates would need to be modified if the radiating
material was moving at relativistic velocities and emitting
non-isotropically.

The millimetre activity linked to the rapid flares is also very
violent. The infrared variations do not formally require that the
emitting electrons have a bulk relativistic motion. The millimetre
emission associated with these flares cannot, however, be understood
in terms of synchrotron cooling by a static source in a constant
magnetic field \cite{Robsonetal93}. Indeed the millimetre activity of
3C~273 is particularly complex with many flares of different
characteristic times. The spectrum of individual flares and their
evolution cannot, therefore, be confidently extracted.

\subsection{The blue bump}
\begin{figure}
  \includegraphics[totalheight=15cm,width=12cm]{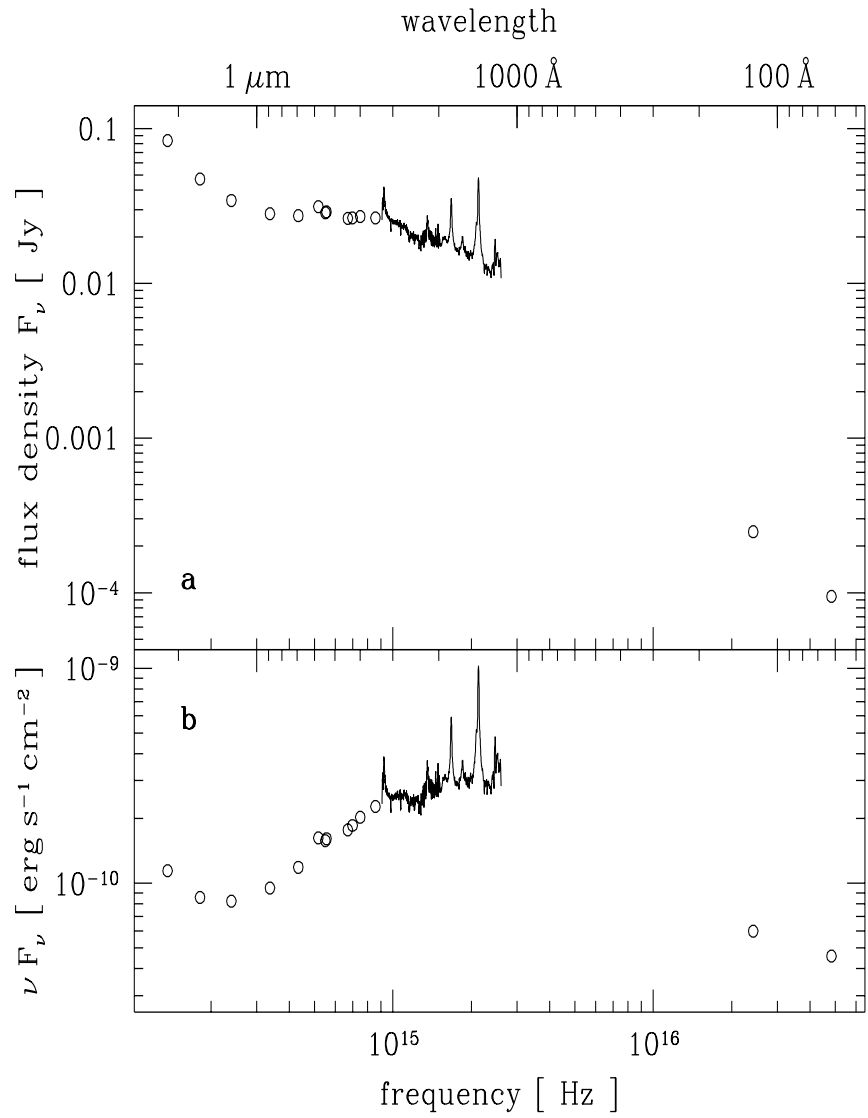} 
\caption{\label{fig:averageoptical}The average optical-ultraviolet spectrum.  Data from 
Fig.~\ref{fig:theflux}. Panels as in Fig.~\ref{fig:average1}.}
\end{figure}

Seyfert galaxies and quasars are generally characterized by an excess
emission in the optical UV region of the spectrum when compared to the
high frequency extrapolation of the infrared continuum. 3C~273 is no
exception. This excess has been called the blue bump or the big blue
bump to distinguish it from a smaller bump due to blended FeII lines.
We will use the first name here. Figure~\ref{fig:averageoptical} gives
the average spectrum obtained in the optical and ultraviolet domains.

It has been proposed by \cite{Shields78} that the blue bump might be
due to thermal emission from the surface of an optically thick but
geometrically thin accretion disc. The structure of these discs in
which the energy released at each radius is the gravitational energy
locally lost by the gas spiraling inward had been calculated by
\cite{ShakuraSunyaev73}. The emitted spectrum is the superposition of
black bodies of temperatures decreasing from the inner radius to the
outer radius of the disc.  \cite{Ulrich81} and \cite{MalkanSargent82}
fitted the observed blue bump emission with this model which they
approximated by a single temperature blackbody. The temperatures they
deduced from the fits were 21\,000\,K and 26\,000\,K respectively. The
size of the source was estimated from the Stephan-Boltzmann law to be
close to $10^{16}$\,cm using the observed flux and assuming as we do
here $H_0 = 50$\,km/sMpc and isotropic emission.  More detailed
attempts to represent the optical-ultraviolet emission of 3C~273 using
standard accretion disc models followed.

The blue bump is broader than the spectrum of a single black body.
Accretion discs do have a distribution of temperatures. Fits to the
data therefore improved when disc models replaced single temperature
representations \cite{Malkan83}. The accretion disc models can be
parametrized in terms of the central mass of the black hole and the
accretion rate. \cite{Malkan83} obtained from his fit to the 3C~273
optical and UV data a mass between 2 and 5 $10^8$ solar masses for the
mass of the black hole and an accretion rate $\dot{M}$ between 4 and
12 solar masses per year. The resulting ratio of the luminosity to the
Eddington luminosity (the luminosity for which gravitational
attraction and radiation pressure balance each other) was slightly
larger than~1. It must be noted that these fits depend on the
assumptions made for the underlying components. Different
extrapolations of the infrared flux onto which the blue bump is added
lead to different blue bump spectral energy distributions to be fitted
by the models \cite{CamenzindCourvoisier84}.

These early papers were followed by a large effort in which different
assumptions were made to describe the components other than the disc
emission (in particular the possibly underlying extrapolation of the
infrared component and the free bound emission were taken into
account). Meanwhile the accretion disc models grew more complete,
including optically thin parts and a corona. These models can account
for many features of the blue bump emission of quasars and are a
natural consequence of accretion of matter with some angular
momentum provided that an adequate source of viscosity is available to
transport angular momentum towards the outer regions of the nucleus.
\cite{Czerny94} provides a review of the arguments in favour of these
models.

There are, however, a number of difficulties with the accretion disc
model in the case of 3C~273 and other well studied Seyfert galaxies
\cite{CourvoisierClavel91}. These difficulties relate to the shape of
the continuum emission, the dependence of this shape on the luminosity
of the objects and the variations observed in the blue bump emission
(see below).

\subsubsection{The variability of the blue bump}

There exists a very long history of observations of 3C~273 beginning
in 1887. The object is indeed bright enough to be measurable on a
large number of photographic plates. The data up to 1980 have been
collected, homogenised and analysed by \cite{AngioneSmith85}. This
light curve shows variations by more than one magnitude and no
strictly periodical signal.

The optical-ultraviolet emission of 3C~273 varies on many timescales.
One form of variation, that due to the synchrotron flares, has already
been mentioned when the infrared variations were discussed. The
synchrotron flares are indeed observed at higher frequencies than the
near infrared into the optical domain \cite{Courvoisieretal88}.
Outside of the periods of intense flaring the contribution of the
synchrotron emission to the blue bump is negligible. This can be
deduced from the fact observed by \cite{Robsonetal86} that the near
infrared emission is not affected when the synchrotron flux decreases.
The synchrotron power law beeing steeper than the blue dump spectrum
will contribute less to the blue bump than to the near infrared
emission. Since its contribution is not measured in the near infrared
it will therefore indeed be negligible compared to the other
components making the blue bump.

Ultraviolet variability of 3C~273 was first discussed by
\cite{CourvoisierUlrich85}. This discussion was expanded using 9 years
of IUE data in \cite{Ulrichetal88}. This work showed that it is not
possible to account for the changes in the continuum spectral energy
distribution by a variable uniform absorbing medium. Indeed such a
medium would have to alter its reddening law (hence its composition)
at the different epochs at which the flux varied. Difference spectra
showed that the variations are more pronounced at short wavelengths
and could be accounted for by a black body of 3-6 $10^4$\,K that changes
its emitting area. The recent data described below suggests more
complex interpretations for the optical-ultraviolet variations.

The blue bump variability can now be well described using the 10 years
of intense monitoring at optical and UV bands obtained since 1985 and
shown in Figs.~\ref{fig:theflux} and~\ref{fig:Lightcurves2}. Analysis
of these data \cite{Paltani95} and \cite{Paltanietal98} in terms of
a structure function shows that the longest timescale on which the
source varies is slightly shorter than a year at the shortest
wavelengths available with IUE (1\,200\,\AA) and longer than 3 years
(i.e.\ longer than a third of the available timespan of controlled
photometric observations) in the V band.

A cross correlation analysis of the light curves shows, furthermore,
that all the light curves are very well correlated at short lags (less
than a month, see below) but that a secondary correlation peak
monotonically increases when longer wavelength light curves are
correlated with the 1200\,\AA\ light curve.

Both of the above observations indicate that the blue bump variations
are of a complex nature and cannot be due to a single physical
component. Indeed, a single component like a black body of variable
emitting area, is expected to show the same variation timescales at
different wavelengths. \cite{PaltaniWalter96} have proposed a
decomposition into two components for a set of AGN including 3C~273
based on the suggestion that one component is stable or at least
varies on timescales much longer than the other and that the spectral
energy distributions of both remain stable, the variability being due
to the changes of the relative normalisation. This decomposition has
the interesting side benefit of giving a very high signal to noise
spectrum that allows the measurement of the reddening, in the case of
3C~273 $E_{B-V}=0.038$, compatible with galactic reddening.

At very short lags, the lag of the peak of the cross correlation
between the 1200\,\AA\ light curve with those at longer wavelengths
increases with wavelength. The lag is of 2 days at 2\,000\,\AA\ and 10
days for the V band \cite{Paltanietal98}. These lags, although
probably significantly different from zero (note that it is difficult
to give a formal uncertainty on the lag at which a cross correlation
peaks) is many orders of magnitudes less than that expected from
viscously heated accretion disc models \cite{CourvoisierClavel91}.
This result is insensitive to the details of the models and is also
valid for other temperature distributions than those of standard
accretion discs. In particular the lag is much shorter than the sound
travel time in the accretion disc between the hot regions emitting the
UV flux and the cooler regions emitting the V band flux. The lag
between the UV and optical light curves of a few days implies that if
an accretion disc is present it must be heated by an external source
rather than by the internal dissipation of gravitational energy in an
optically thick medium. More generally, this result states that the
causal connection between the hot and cool regions that form the blue
bump (i.e.\ those emitting in the ultraviolet and those emitting in the
visible) must be based on information transported at speeds close to
that of light. This observation is similar to that obtained for those
Seyfert galaxies for which adequate data have been obtained. 

The fact that the energy source should be located outside the disc has
several implications. First the standard disc structure and spectra as
deduced using the local gravitational energy dissipation
\cite{ShakuraSunyaev73} is not applicable, secondly the origin of the
heating source must be sought.  In other words, one should find a way
of radiating the energy freed by the accretion process outside the
disc.

One possibility that has been studied is the presence of hot coronae
surrounding the discs \cite{Haardtetal94}. In this paper Haardt et
al.\ consider a structured corona in which a fraction of the accretion
power is released through magnetic interactions. The hot blobs in the
corona reprocess a fraction of the disc soft photons to X-rays.

\subsection{X and gamma-ray emission}

\cite{Bowyeretal70} have reported the first convincing evidence for
X-ray emission from 3C~273 using a collimator instrument on a sounding
rocket. This result was confirmed by Uhuru measurements reported by
\cite{kellogetal71}. In a 1977 review \cite{GurskySchwartz77} state
that 3C~273 is still the only quasar reliably associated with an X-ray
source and that it is not certain that X-ray emission is a
characteristic of active galactic nuclei in general. This has changed
since then, X-ray emission is one of the important emission components
of all classes of AGN. 3C~273 has thus been observed, often many
times, by all X-ray satellites. We present here the X-ray data to
about 10 keV obtained by \textsc{Einstein} \cite{Wilkesetal87}, EXOSAT,
GINGA (both reported in \cite{Turneretal90}), ROSAT
\cite{Leachetal95} and \cite{Walteretal94}, ASCA \cite{Yaqoobetal94}
and SAX \cite{Grandietal97} and higher energy data as discussed in
\cite{Maisacketal92} for HEXE data, \cite{Bassanietal92} for SIGMA
data and \cite{Mcnaronetal95} for OSSE data.

\begin{figure}
  \includegraphics[totalheight=15cm,width=12cm]{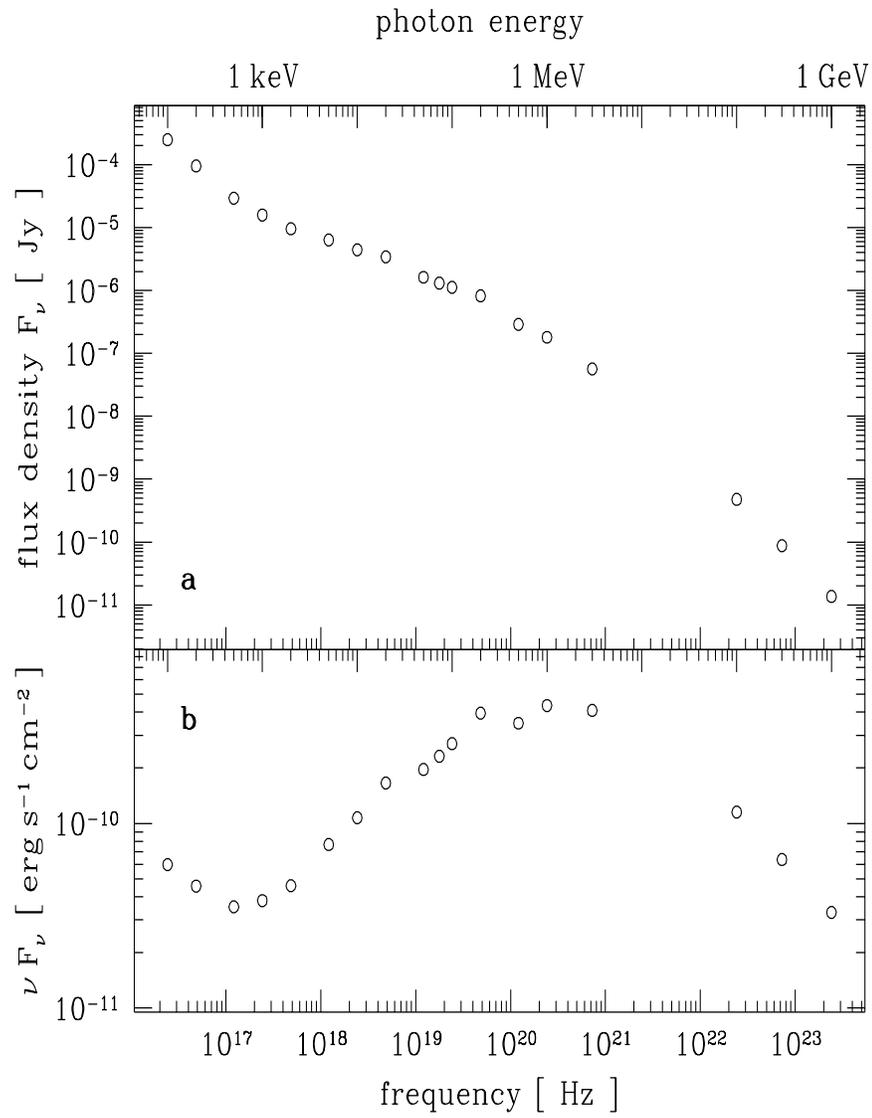}
\caption{\label{fig:averagex}The average X and gamma-ray spectrum. Data from Fig.~\ref{fig:theflux}. Panels as in Fig.~\ref{fig:average1}.}
\end{figure}

This emission has 4 features (see Fig.~\ref{fig:averagex}): A steep low energy
component that emerges from the interstellar absorption called the
soft excess, a straight power law that extends to about 1MeV (which we
will call the medium energy component) on which a weak Fe line appears
and a steeper power law above about 1\,MeV (called the high energy
component in the following).

\subsubsection{The soft excess}

Low energy X-ray data on many AGN cannot be fitted by extrapolation of
the power law that is used to describe the medium energy component.
The flux observed at low energies ($\lesssim 0.5$\,keV) is
systematically larger than that predicted by this extrapolation. Since
the presence of cold matter along the line of sight decreases the low
energy X-ray flux we can deduce that there are no large amounts of cold
material in excess of that within our Galaxy. The soft excess (common
also in Seyfert galaxies) is poorly characterized as it extends into
the low energy region in which the flux is completely absorbed by
interstellar matter, because only few energy resolution elements are
available (this is expected to change with the launch of the XMM
satellite in 1999) and because the amount of absorbing matter on the
line of sight is not known. One way of fitting the data is by an
optically thin thermal emission model of $k\cdot T \simeq 0.2\,$keV,
another is to use a power law of photon index -2.7
\cite{Leachetal95}. These fits should be taken as a mathematical
description of the shape of the emission rather than as a true
physical description of the emission.

Simultaneous observations in the UV and X-ray domains in a set of
objects including 3C~273 \cite{Walteretal94} and \cite{Walteretal93}
indicate that the parameters of the soft excess and those of the blue
bump are correlated. Should this be confirmed, it would show that the
soft excess is the high energy tail of the blue bump. This high energy
end of the blue bump would then occur at roughly the same energy for
nuclei of very different luminosities, a fact unexpected in standard
accretion disc models in which the maximum temperature is expected to
decrease with the 1/4 power of the luminosity
\cite{CourvoisierClavel91}.

\subsubsection{The medium energy component}

At energies higher than about 0.5\,keV, the spectrum is well described
by a single power law extending to the MeV region. There is no hint of
any curvature in this spectrum, contrary to predictions of models in
which the X-ray flux is in part reflected by the surface of a cool
disc. These latter models predict the presence of a so-called
reflection hump corresponding to the Compton reflection of the primary
component. This hump is observed in many Seyfert galaxies
\cite{Mushotzkyetal93}, but not in 3C~273 \cite{Maisacketal92},
\cite{Grandietal97}. This result is surprising because 3C~273 has a
bright blue bump. Indeed, reprocessing models (models in which the
blue bump emission is due to a disc heated by an external X-ray
source) predict that an important blue bump would be linked to the
presence of re-processing signatures also in the X-ray domain.

The spectral slope of this component is typically 0.5 and shows
evidence for some variations (see below). As a result, the energy
radiated per logarithmic energy interval ($\nu \cdot f_{\nu}$) peaks
at the energy at which the spectral break is observed, i.e.\ around
1\,MeV.

\subsubsection{The Fe line}

A further signature of X-ray reprocessing by cold material is the
presence of a fluorescence emission line at 6.4\,keV. There is some
evidence for the presence of a weak line at this energy in the X-ray
spectrum of 3C~273. One of the GINGA observations (in July 1987)
showed evidence for the line at the 99\% confidence level. The line
equivalent width was 50\,eV, the corresponding line flux was $4.5 \pm
2.5\, 10^{-5}$\,photons\,cm$^{-2}$ s$^{-1}$ \cite{Turneretal90}. The
other GINGA observations reported by \cite{Turneretal90} provided only
upper limits to the line flux compatible with the one of July 1987.
Since the continuum flux had varied significantly this means that the
line equivalent width did vary. A further line detection is reported
by \cite{Grandietal97} in a SAX observation in July 1996. The
equivalent width was $30 \pm 12$ eV.  Other observations, by EXOSAT
and ASCA provided only upper limits \cite{Turneretal90} and
\cite{Yaqoobetal94}. This somewhat confused set of measurements
probably means that there is a weak line present in the emission of
the quasar but that the data available do not allow us to perform a
convincing analysis of the line variability nor of its relationship
with the continuum variations. This study will be of prime importance
to understand where the cold matter emitting the fluorescence line is
located with respect to the primary X-ray source.

\subsubsection{The high energy component}

Early gamma ray detections were due to the Cos-B satellite. The source
was identified in the data from the position coincidence of a compact
source (i.e.\ a source not resolved by Cos-B) with the position of the
quasar \cite{Swanenburgetal78}, \cite{Bignamietal81}.

The flux above a few MeV is well described by a power law of index
$1.4 \pm 0.1$ \cite{Lichtietal95}. This index has been shown to vary
between $1.2 \pm 0.2$ and $2.2 \pm 0.5$ \cite{vonMontetal97},
hardening with increasing flux (we give here energy spectral indices
rather than photon indices to remain consequent with the discussion of
the spectrum at lower energies). Quite expectedly, the spectral index
around 1\,MeV is between the X-ray spectral index and the one observed
at higher energies as it is in this region that the spectrum steepens
from a slope of about 0.5 to to one of 1.5. It must be stressed that
the high energy component described here is a power law and not an
exponential cut-off of the medium energy component. This spectral
break is an important constraint to any model, it is larger than 0.5,
the value expected from simple one component Compton cooling models.
Models for the high gamma-ray emission of beamed AGN include the
relativistic electron positron beam model of \cite{Marcowithetal95}.
In this model the emerging emission is due to inverse Compton process
of relativistic electron position pairs on the soft photons from the
accretion disc. The observed spectral break is due to the energy
dependence of the gamma-ray photosphere defined by the optical depth to
pair production being equal to one. A further model is suggested
by \cite{Mannheim94} and \cite{Mannheim93} in which ultra relativistic
protons generate gamma ray photons via pion and pair photo production.

Several candidate models have been fitted to the data by
\cite{Lichtietal95} and \cite{vonMontetal97}. They are all based on
the assumption that the gamma ray component is emitted by electrons
or/and hadrons in the relativistic jet. This assumption is due to the
remark that the high energy photon density estimated from the observed
flux and the time scale of variability implies that the
electron-positron pair production optical depth is considerably larger
than one. This would imply that the high energy photons cannot escape
from the source region. This is formally described by the
dimensionless compactness parameter $l$:

$$l =\frac{\sigma \cdot L}{mc^3 \cdot R}\,,$$

where $\sigma$ is the pair production cross section, L the luminosity
and R the size of the source as deduced from the source variability
timescale. Using $\sigma \simeq \sigma_{Thompson}$, a variability time
scale of 0.5 days as given by the Ariel V measurement described below
and a maximum observed gamma ray luminosity, \cite{Lichtietal95} deduce
a compactness of 210, implying that the optical depth of the region to
pair production ($=l/4\pi$) is much larger than~1. Using a more
established variability time scale for the medium energy component of
several days does relieve the compactness question and lessen the
justification for identifying the high energy emission of 3C~273 with
the relativistic jet. It should be noted, however, that in other
sources, the BL Lacs observed at very high energy, the compactness is
such that the gamma ray emission must be emitted by strongly
relativistic jets, justifying a similar assumption also in the case of
3C~273.

\subsubsection{X-ray variability}

The X-ray flux of 3C~273 varies considerably. The first report of this
variability is by \cite{Marshalletal81} who detected an 80\% flux
increase in a 40\,000s Ariel V observation. This is the only report of
very strong and short flux variation in 3C~273. Since Ariel V was not
an imaging instrument, there is a possibility of source confusion.
This result is therefore in strong need of confirmation by an imaging
instrument.

The flux of the soft excess has been found \cite{Courvoisieretal87} to
be variable in time and to vary independently of the medium energy
component.  \cite{Courvoisieretal90} reported a variability of 4\% in
about 17 hours in the medium energy component observed by EXOSAT.
\cite{Leachetal95} analysed 14 ROSAT observations. They confirm that
the soft and medium energy X-ray components vary independently and
reported their fastest variations to be by about 20\% in 2 days in the
soft energy band (0.1-2.4 keV) observed by ROSAT.

The 2-10\,keV flux observed by EXOSAT and GINGA indicate variations
from $0.60 \cdot 10^{-10}$ to $ 1.68 \cdot 10^{-10}$\,ergs cm$^{-2}$
s$^{-1}$ for the medium energy component. This was obtained with 13
observations spanning about 5 years. The resulting light curve is
hopelessly undersampled, preventing any description of the flux
variations.

This data set also provided evidence that the spectral slope in the
same energy domain varies by small but significant amounts. The
spectral slope is not correlated with the flux but is anti-correlated
with the logarithm of the 2-10\,keV count rate divided by the UV photon
rate \cite{WalterandC92}.  This result is interpreted in this paper as
being the signature that the medium energy X-ray component is due to a
thermal Comptonisation process of the UV photons by an electron
population of about 1\,MeV covering a few percent of the UV source with
an optical depth of 10-20\%. Work in progress shows that when using
the BATSE data as displayed in Fig.~\ref{fig:lightcurves} one finds a weaker
anti-correlation between the spectral slope. This work should,
however,take into account the delay between the UV and the X-ray light
curves. This delay (discussed below) indicates that the UV flux
Comptonised in the hot regions may not be that observed
simultaneously, but rather the UV flux observed about 2 years earlier.
This may not have had much effect in the early phases of the
monitoring because the UV flux was relatively quiet compared with
later epochs.

\subsection{The Overall continuum spectral energy distribution\label{sect:3.5}}

\begin{figure}
  \includegraphics[totalheight=15cm,width=12cm]{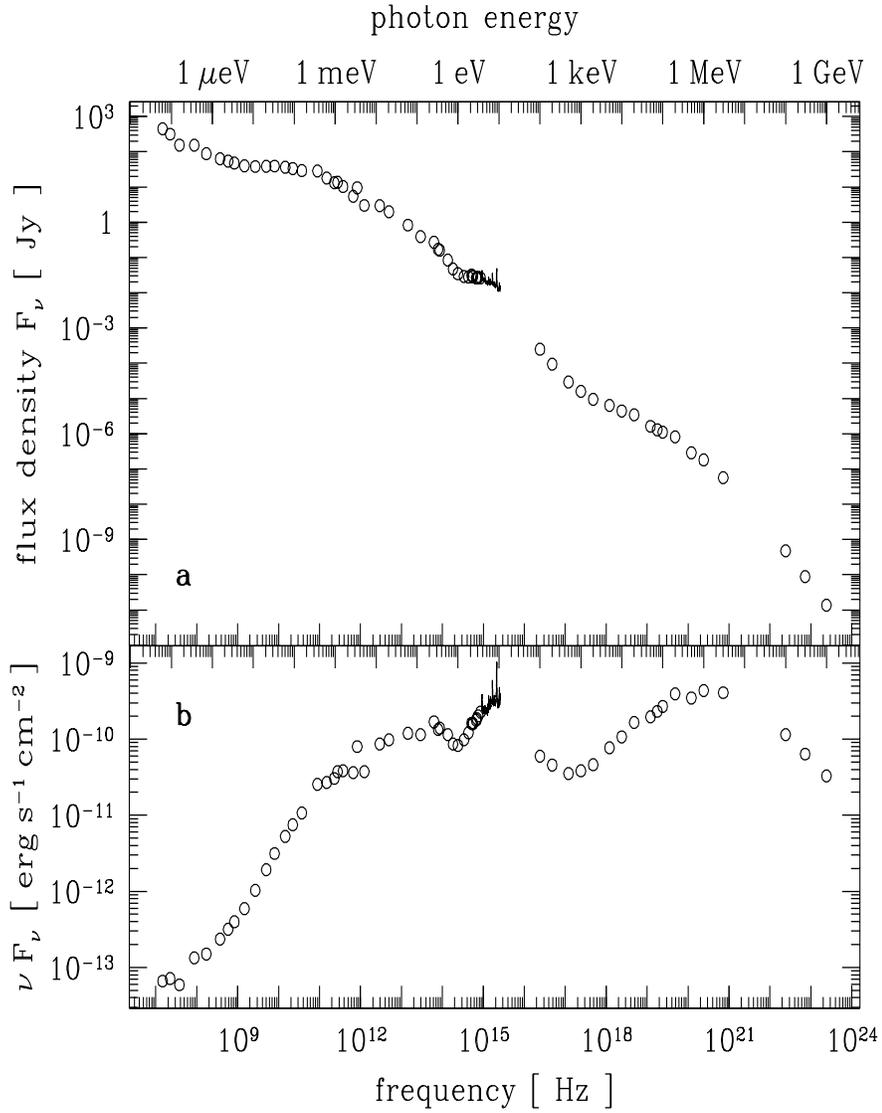}
\caption{\label{fig:Overallspectrum}Overall average spectrum of 3C~273 (first panel). This corresponds to a projection onto the frequency axis of all data in figure 1. The bottom  panel shows the same data as above but represented as  $\nu\cdot f_{\nu}$ versus $\nu$.}
\end{figure}

Having studied the emission components separately we can now put all
these elements together. To do so we present in
Fig.~\ref{fig:Overallspectrum} the average spectrum obtained by
projecting all the data of Fig.~\ref{fig:theflux} onto the frequency axis.
Figure~\ref{fig:Overallspectrum} also gives the same data but in the
form of $\nu\cdot f_{\nu}$ versus $\nu$. It is striking that the flux
per logarithmic interval is nearly constant over more than ten decades
of frequency, another way of expressing that to the first order the
emission is proportional to $\nu^{-1}$. In the second order, it is
striking to see two maxima in the $\nu\cdot f_{\nu}$ versus $\nu$
distribution at roughly the same level, one in the far ultraviolet and
the other at about 1\,MeV.

Integrating the spectrum one can deduce the total flux in the average
spectrum and the bolometric luminosity of 3C 273. One finds a total
flux of $1.9 \cdot 10^{-9}$ ergs s$^{-1}$\,cm$^{-2}$ and assuming
isotropic emission, $H_0 = 50$\,km/(s Mpc), $\Omega = 1$ and $q_0 =
0.5$ one finds a luminosity of $2.2 \cdot 10^{47}$\,ergs s$^{-1}$.
(T³rler et al in preparation).

\section{The multi-wavelength approach to the continuum emission}

Studying each component or equivalently the observations in individual
spectral domains provides information on the physical processes at the
origin of the components. It does not, however, provide a full picture
in which it is possible to see how the gravitational energy released
by accretion is distributed between the various cooling channels, nor
does it allow us to describe the respective geometrical arrangements
of the components or their physical relationships. Correlated studies
across the complete electromagnetic spectrum are necessary for this
research. 3C~273 is a prime source for these studies as it is bright
in all bands and located close to the celestial equator. Both
characteristics provide for ease of access with many instruments.

An early and surprising result was obtained by
\cite{Courvoisieretal90} who showed that the UV light curve leads the
radio emission by a few months. This result is confirmed by continued
monitoring \cite{Courvoisier97} and may be understood if the blue band
flux is a signature of the accretion process (and hence of the energy
release) and the radio (synchrotron) emission one of the cooling
channels located at some distance from the central black hole. In this
case and assuming that part of the accretion energy is carried along
the VLBI jet (described in Sect.~\ref{subsec:VLBI_Jet}) and with its
velocity from the central source to the location of the radio
emission, the observed delay of approximately 0.4 year can be used to
estimate the distance D at which the radio (22\,GHz) flux is emitted:

$$D = c \cdot \Delta t \cdot (\frac{1}{\beta_j} - \cos \theta_j)^{-1}
, $$

where $\beta_j = 0.95$ is the VLBI jet velocity divided by $c$ and
$\cos \theta_j = 0.95$ the cosine of the jet angle with respect to the
line of sight \cite{Davisetal91}. These data imply that the radio
emission is located some 4 light years from the central source along
the jet.

\cite{Clementsetal95} have also performed a correlation analysis of
the blue bump and radio light curves. They used photographic
photometry from the Rosemarie Hill observatory taken from 1974 to
1992, radio data from the University of Michigan Radio Astronomy
Observatory and data from the Algonquin Radio Observatory from 1966 to
1990. Cross correlating the optical and radio light curves does not provide a
significant signal at any lag. This stems most probably from three
factors. The blue bump light curve is undersampled and was obtained
from B band observations rather than ultraviolet. Furthermore the
radio light curve is at 10\,GHz, less than that used in the preceding
analysis. At lower radio frequencies, the light curves are smoother
and the amplitude of the variations decreases \cite{Terõsrantaetal92}.
This effect probably smoothes the variations in 3C~273 to the point at
which the correlation with the optical variations is lost. It is
interesting to note, however, that \cite{Clementsetal95} and
\cite{Tornikoskietal94} do find significant correlations between the
radio and the optical light curves in several other objects. The
optical light curves always lead the radio light curves. Typical lags
are of the order of several months similar to those obtained in the
case of 3C 273.

Using the radio and ultraviolet data, one may also wonder whether it
is possible that 3C~273 is a mis-directed BL Lac object. Indeed 3C~273
has a strong synchrotron source emitting in the radio and millimetre
domain and a superluminal jet. Both are characteristic of BL Lac
objects. Were the blue bump and the emission lines overwhelmed by the
synchrotron emission, one might well classify 3C~273 as a BL Lac type
object. For this to be the case, the synchrotron component should,
however, be boosted by a factor larger than $10^3$
\cite{Courvoisier88}. The resulting radio flux would then be larger
than $3 \cdot 10^4$\,Jy, a highly improbable flux. It would thus appear that
the presence of a strong blue bump and bright emission lines is an
intrinsic difference between BL Lac objects and quasars like 3C~273
rather than due to orientation effects.

The relation of the UV with the X-ray emission is of considerable
interest to test reprocessing models. One aspect of this correlation
has already been discussed when the slope of the X-ray component was
compared to the ratio of X-ray to UV photons. Cross correlating the
IUE and BATSE light curves, one finds a very significant correlation
peak at a lag indicating that the X-ray light curve follows the UV
light curve by 1.75 years and no significant correlation close to zero
lag \cite{Paltanietal98}. Assuming that this result represents a
physical reality rather than a chance occurrence of features in the
light curves, \cite{Paltanietal98} conclude that the Comptonising
X-ray emitting medium could be heated in a shocked region formed in a
mildly relativistic wind at about 1\,pc from the central source, in good
agreement with the model proposed by \cite{CC89} and described in
Sect.~\ref{sect:8}. It is, however, also possible to apply models in
which the Comptonising medium is located on the surface of the soft
photon source (e.g.\ on the surface of an accretion disc). In this
case, the flux correlation cannot have a physical meaning and,
provided that the temperature of the plasma is known (e.g.\ from
\cite{WalterandC92}), one deduces an X-ray spectral slope as a
function of time assuming a variable optical depth in reasonable
agreement with the existing data.

Looking at the correlations between the X-ray and the radio light
curves, one finds no significant correlation at zero lag
\cite{Courvoisieretal90}, \cite{Courvoisier98}.  This indicates that
the X-rays cannot be due to a simple synchrotron self-Compton process
in which the radio photons are scattered by the same electron
population that produced them in the first place. The light curves now
available indicate that the medium energy X-ray and radio components
are correlated when the X-rays emission follows the radio by 2.2
years. This result is however, based on a dominant flare in the X-ray
light curve, it remains then to be seen whether it proves solid with
time.

Taken at face value the data presently available indicate that the UV
flux leads all the other components. The typical delays are of the
order of one or very few years. This result is in strong need of
confirmation. It will, however, take many years of careful
multi-wavelength observations to do so. This result would considerably
strengthen the conjecture that the UV emission is a signature of the
accretion and that the released energy is transported from the central
regions of the gravitational potential well to the regions where it is
radiated by relativistic or near relativistic flows.

\section{Line Emission}

Line emission is one of the defining properties of quasars. It has
also been one of the main lines of research over the last three
decades. The information content of the spectrum is -trivially- richer
in the lines than in the continuum. Unfortunately, the information
gained from the emission line does not allow researchers to gain much
understanding on the mechanisms at the origin of the radiation. This
must be obtained out of the relatively information poor continuum. What
the information provided by the lines allows us to do is to describe
the gas surrounding the energy source. This includes a description of
the physical state of the gas (temperature, density, ionization
level), of its kinematics through the width of lines and of its
geometrical arrangement (filling factor) through the equivalent width
of the lines. A general review of these inferences can be found in
\cite{Peterson97} and \cite{Netzer90}. A detailed fit to all line
features in 3C~273 can be found in \cite{Willsetal85}.

The classical picture is that the lines are formed in a set of
photoionized "clouds" in rapid movement around the central black hole.
The smoothness of the lines implies that the number of clouds must be
large. How large is however not known yet (see Dietrich et al.\ in
preparation). The continuum at the origin of the photoionization is
normally associated with the central source.

Whereas one would expect that the study of the line intensity ratios
should be able to provide a description of the ionizing continuum
(since the lines come from elements that have different ionizing
potentials) particularly in the unobservable part of the spectrum
between 912\,\AA~and $\sim$0.1\,keV few concrete results have emerged
\cite{Binnetteetal88}, \cite{Kroliketal88}.

\cite{Willsetal85} claim that the photoionization models they use
represent well all the line features of the quasars they describe with
the notable exception of the FeII line blends observed in the optical
and UV parts of the spectrum. 3C~273 is no exception to this "FeII
problem". Expected values for the intensity ratios of FeII lines to
Ly${\alpha}$ is of 0.3-0.5 \cite{Netzer90} whereas the observed ratio
(corrected for a reddening A$_V = 0.16$ corresponding to the galactic
reddening E$_{(B-V)} = 0.05$ as used in \cite{Ulrichetal88}) is
slightly larger than 1 \cite{Willsetal85}. Clearly, assuming a more
important intrinsic reddening will tilt the FeII lines to Ly${\alpha}$
ratio to smaller values, lessening the problem; there is, however no
reason to assume a large intrinsic reddening (see above).  This
problem is as of yet unsolved and may point to additional energy
sources in the broad line clouds (e.g.\  mechanical heating) and/or to
a more complex structure of the broad line region than envisaged in
most photoionization calculations \cite{CollinSouffrinLasota88}.

Whatever the details of the fits and the agreement of the line
intensities with various photoionization models, a very important
point is that the heavy element abundances are large, in some
instances larger than solar. This indicates that the gas surrounding
quasars, and in particular 3C~273, has been going through one or
several generations of star formation and explosion before being found
in the very inner regions of the galaxies hosting the quasar.

\subsection{Line variability}

Some carefully designed observation campaigns using the IUE satellite
and additional ground based data have shown that the emission line
variations in Seyfert galaxies follow those of the continuum. The lag
is short and increases as the level of ionization decreases. The
amplitude of the variations in the high ionization lines and in
particular of Ly${\alpha}$ are similar to those of the continuum.  See
\cite{Peterson93} and references therein for a review of these
results. This is one of the strongest arguments demonstrating that the
emission lines are indeed due to photoionization of gas close to the
continuum source.

In 3C~273, the picture is quite different. Whereas the UV continuum
varies by a factor of about 2 (see above), the amplitude of the
Ly${\alpha}$ variations is only 15\% or less \cite{Ulrichetal93},
\cite{Ulrichetal88}, \cite{OBrienetal89}. \cite{OBrienetal89} studied
the timescale of the Ly${\alpha}$ variations and the possible
Ly${\alpha}$ continuum correlations. They claimed that the observed
timescales are less than one year and that there is some correlation
between line and continuum variations. The existence of such
correlations and the measurement of any lag between continuum and line
light curves using variations as small as those observed in 3C~273
are, however, barely possible based on the IUE data base
\cite{Ulrichetal93}.
The small amplitude of the line variations compared with the continuum
variations is confirmed by a study of the IUE data on 3C~273 up to
1991 by \cite{TurlerandCourvoisier97} which shows that when performing
a principal component analysis of the spectra of 3C~273, the principal
component does not show a line, but only the continuum. This analysis,
contrary to previous ones considered all the spectra of a single
object as the matrix in which the principal component is to be sought.
The principal component then gives for the given object the most
variable "spectrum". The result obtained for 3C~273 is in contrast
with other well studied objects for which the principal component has
the same shape as the average broad line. In the case of 3C 273 the
principal component is essentially flat, indicating that the continuum
varies, not the lines. This is possibly due to the fact that in
3C~273, an intrinsically bright object, the broad line region is
further from the central source than the characteristic time of the
continuum variations (of the order of a year) times the velocity of
light.

\section{The Jet}

There are two very different aspects to the jet in 3C~273. One is the
small scale jet observed with VLBI techniques and showing superluminal
motions. The other is the long jet visible at radio, optical and X-ray
photon energies from $12\prime\prime$ to $22\prime\prime$ from the
core at a position angle of 222$^{\circ}$.

\subsection{The radio-optical jet}

It was already clear at the time of the identification of 3C~273 with
an optical object that the radio source had two components called A
and B \cite{Haz63} and that the optical counterpart was coincident
with the B component. The optical counterpart had some luminosity in
the form of a jet extending in the direction of the A component and
ending precisely at its position, \cite{Schmidt63}. This established
from the earliest time that 3C~273 has a one-sided jet. No counterjet
has ever been discovered either at radio or at optical wavelengths.

Modern MERLIN and VLA maps and polarization have been presented in
\cite{Conwayetal93}. At radio frequencies, the jet flux increases
monotonically from the inner radius to a peak located at roughly
$21^{\prime\prime}$ from the core and then falls steeply. The
behaviour is very similar for wavelengths between 2\,cm and 73\,cm.
The polarization at 6\,cm is such that the B-field is parallel to the
jet axis except at the point of maximum flux (the so called hot spot)
where it is perpendicular to the jet axis. At other wavelengths the
polarization is qualitatively similar. The maximum polarization is
approximately 20\%\, with three spots along the ridge at $14^{\prime\prime}$, $17.5^{\prime\prime}$
and $20.3^{\prime\prime}$ from the core where the level of polarization is very low.

The optical jet is more structured than the radio jet, it is a
succession of bright spots and regions of weaker emission. Most modern
work on the jet uses a simple denomination for the hot spots given by
\cite{Lelievreetal84}. The spot nearest to the core is labeled A, the
subsequent spots are B, C and D. Subsequent work at higher angular
resolution then subdivides the spots in sub-units.
Figs.~\ref{fig:subfiguresjet1} and~\ref{fig:subfiguresjet2} give an
optical image of the jet and labels the main spots.

Recent optical data on the jet of 3C~273 and a detailed comparison
with the radio morphology and polarization of the jet are presented in
\cite{RoeserMeisenheimer91}, \cite{Roeseretal96}, \cite{Bahcalletal95}
and in \cite{Roeseretal97}. It results from these studies that the
position angle of the jet as observed in the radio and optical domains
is the same at $222.2^{\circ}$. The general appearance of the jet in the
2 spectral domains is similar at first sight. There are, however, some
notable differences. The radio jet can be followed from the core of
the quasar to the brightest regions, whereas the optical jet can only
be observed from some distance to the core onward. This statement is,
however, clearly of limited value as it results from systematic
observation limitations in detecting weak surface brightness features
close to a bright point source. It is nonetheless clear that the ratio
of optical to radio flux changes with distance to the core. The high
resolution of the HST images reveal that the optical jet is structured
on smaller scales than the radio jet and that it is narrower. The
optical images show two "extensions" outside the jet axis at the inner
and outer edges. At least the outer extension seems to be due to the
presence of a spiral galaxy on the line of sight \cite{Roeseretal97}.

\sbox{\MaBoiteA}{\cite{Bahcalletal95}}
\begin{figure}
\includegraphics[scale=0.77]{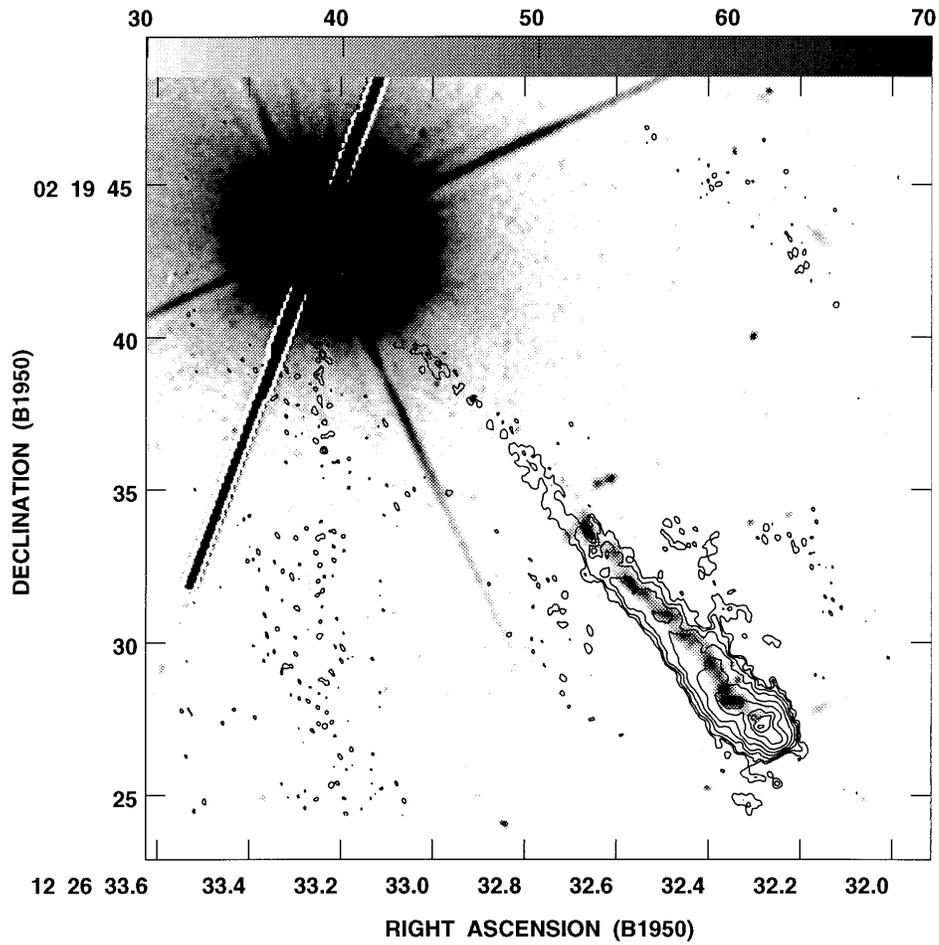}
  \caption{\label{fig:subfiguresjet1}The core and jet of 3C~273 as
   observed in the radio domain (contours) and with HST (grey scale).
    This shows that while the radio jet can be followed all the way to
    the core, the optical jet can only be detected in the outer
    regions of the jet. This figure is from \usebox{\MaBoiteA}.}
\end{figure}

\sbox{\MaBoiteA}{\cite{Bahcalletal95}}
\sbox{\MaBoiteB}{\cite{Lelievreetal84}}
\begin{figure}    
      \includegraphics[scale=0.67]{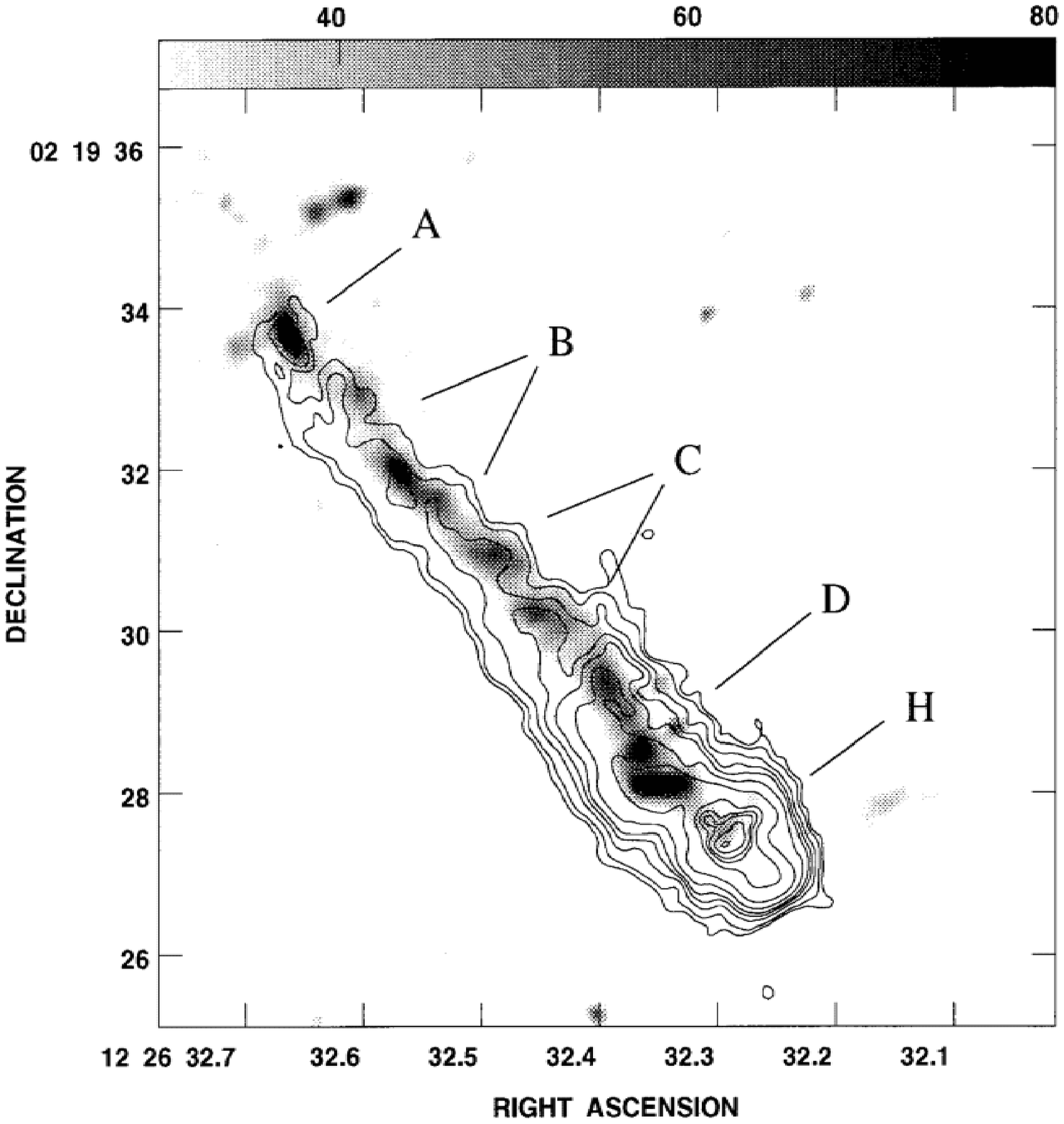}
     \caption{\label{fig:subfiguresjet2}Same data but showing an 
       enlargement of the jet and the nomenclature of
       \usebox{\MaBoiteB}. This figure is from~\usebox{\MaBoiteA}.}
\end{figure}

The radio and optical data may be combined to obtain spectral energy
distributions for the knots of the jet.The innermost knot (A) has a
straight power law continuum without any sign of a cut-off in the
visible or UV, whereas the subsequent knots are well described by a
power law extending from 100\,MHz to $10^{14}$\,Hz followed by an
exponential cut-off.

The optical and radio polarization of the jet are also similar in
their main properties and also show some discrepancies when looked at
in greater details. The maximum polarization is in both cases of the
order of 10\%, low in the inner regions ($< 15^{\prime\prime}$) and
rising towards the hot spot.

Spectral energy distributions and the relatively high degree of
polarization both suggest that the jet emission is due to synchrotron
processes. Synchrotron cooling of electrons depends on the inverse
square root of the frequency. Optically emitting electrons are
therefore expected to be associated with the recent history of the
electron acceleration, while radio emitting electrons remember a
history thousand times longer. It is thus not necessarily expected
that the radio and optical jets should be similar in the location of
the hot spots. It should be noted, however, that the presence of bulk
relativistic motions would tend to enhance for any observer the
emission of those parts of a jet that move in directions close to the
line of sight. Such relativistic effects are independent of the
frequency and would lead to the presence of coincident bright spots in
both the radio and optical images. \cite{Bahcalletal95} suggest one
model along these lines in which the jet structure would be due to a
helical bulk motion of the emitting electrons within the jet.

Radio profiles of the jet perpendicular to the jet axis
\cite{Roeseretal96} show that the jet is symmetric at large distances
from the core, whereas it is more extended in the South closer to the
core (at $15^{\prime\prime}$). This extension is not observed in the optical domain
and has a steep spectral index. It is suggested that this is emission
from material that has passed through the terminal shock and flows
backwards along the jet.

Extended X-ray emission in the vicinity of 3C~273 has been detected
with the EINSTEIN satellite \cite{Willingale81}. It was found that
there is an excess emission in the approximate direction of the jet at
distances from the core of the quasar that are compatible with the
position of the radio and optical jet. The main difficulty associated with these
data is that the angular distance of the X-ray excess falls well
within the EINSTEIN point spread function. The data were re-analysed
by \cite{HarrisStern87} who deduced a position for the centroid of the
excess at $16^{\prime\prime}$. At this position and with the flux they deduced it was
difficult to interpret the origin of the X-ray flux in terms of
synchrotron, inverse Compton or thermal emission.

\cite{Roeseretal96b} report on a long (17.2\,ksec) ROSAT HRI observation
of 3C~273 in 1992 to which another 68.2\,ksec obtained in very early
1995 were added. These authors used the emission from the core of
3C~273 to center the point spread function every 50\,s and were thus
able to correct for the wobble of the spacecraft. This lead to a point
spread function of $4.5^{\prime\prime}$. Two extended features can be observed in the
resulting X-ray image. One is at position angle 71$^{\circ}$ and is
assumed to be from a weak X-ray source not associated with 3C~273
while the other is at position angle 219.3$^{\circ}$, close to that
observed in the other wave bands. The main contribution comes from a
distance to the core of $15^{\prime\prime}$, similar to what had been obtained with
EINSTEIN\@. The flux derived from the ROSAT observation is, however,
considerably less than that derived from the EINSTEIN data.

It was noted above that the A knot is very blue and showed no evidence
of a cutoff at high frequencies. Should the X-ray emission be indeed
associated with this knot, then the flux would lie on the
extrapolation of the radio-optical spectral energy distribution. The
position of the X-ray excess is, however, such that it does not
coincide with knot A (nor with any of the sub-knots derived at high
angular resolution). The next knot (B) has a spectrum that shows a
clear cut-off in the optical spectrum and is therefore unlikely to
extend to the X-ray domain. The difficulties in the interpretation
raised by \cite{HarrisStern87} remain therefore. Were the X-rays due
to inverse Compton processes, one would like to understand why this is
observed only at this location rather than associated with all the
knots.

The extended jet of 3C~273 is still not completely understood. It is
also noteworthy that this is the only jet from a quasar from which
optical and X-ray emission have been detected to date. Optical emission
from the jets of extra-galactic radio sources is a rare phenomenon, it
is therefore not surprising that no other quasar jet has been observed
at higher frequencies than the radio domain. It will nonetheless make
the task of understanding the nature of the extended jet of quasars
very difficult.

\subsection{The VLBI Jet}\label{subsec:VLBI_Jet}

The core of 3C~273 (3C~273 B as it was called in the 1960s) is a very
bright radio source. The presence of the jet described above at large
angular distances showed that the source is not point like, but has an
interesting geometry. Both facts made 3C~273 a prime source for high
resolution radio observations as these became possible by using
together telescopes spanning approximately the size of the Earth (Very
Long Baseline Interferometry, VLBI in short). Early observations used
few telescopes and did not produce maps but were able to measure
whether or not the source is extended at certain angular scales.
\cite{Brotenetal67} and \cite{Clarketal67} showed thus that 3C~273B
has an angular size less than $0.005^{\prime\prime}$ at a wavelength of 18cm.
Structure on the scale of milli arcseconds (mas in the following) has
been found by \cite{Knightetal71} and \cite{Cohenetal71} who also note
a difference between their results that can be interpreted by a change
in angular size of the source. Radio data (visibility functions rather
than maps) confirmed the reality of the changes and revealed a steady
expansion of the source between 1970 and 1977 \cite{Cohenetal77}.
Study of the location of the minima in the visibility curves showed an
apparent expansion velocity of 5.2\,c ($H_0=55$\,km/(s $\cdot$ \,Mpc))
\cite{Cohenetal79}. Maps were also obtained then, showing for the
first time a real jet structure at a position angle of $-117^{\circ}$,
not aligned with the larger scale jet described above
\cite{Readheadetal79}.  Subsequent maps at higher angular resolutions
and using more antennae than previously, thus improving the image
quality, showed that the local maximum of the jet moves away from the
core. The distance from the core to the main jet feature had increased
from 6\,mas in 1977.5 to 8\,mas in 1980.5 \cite{Pearsonetal81}
corresponding to an apparent expansion velocity of $9.6 \pm 0.5$\,c
($H_0=55$\,km/(s $\cdot$\,Mpc)).

More recent VLBI observations have continued the work done at cm
wavelengths, have used higher frequency observations to increase the
angular resolution and have improved on the dynamical range to study
weaker features. These modern data have confirmed the picture
described above and added several new features.

A set of several VLBI observations in the 1980s has revealed that new
jet components (often called blobs) appear every few years. These
components can be followed from one observation to the next and their
projected trajectories mapped. It is thus possible to trace back each
component to the time of zero separation from the core
\cite{Krichbaumetal90}. One of the component was observed to be thus
"born" shortly after a violent synchrotron outburst that had been
observed at wavelengths as short as the visible band in March 1988
(see above; \cite{Courvoisieretal88}). This close association suggests
that in general new components in the jet follow synchrotron
outbursts. This is indeed claimed in a study of \cite{Abrahametal96}
in which the ejection time of 8 components is computed and 
qualitatively compared to single dish light curves. We show in
Fig.~\ref{fig:mmandinfrared} the high frequency radio light curves
and a near infrared light curve available to us (see above) and the
epochs of appearance of new jet components as computed by
\cite{Abrahametal96}. Whereas it seems clear that the ejection of C9
is associated with the infrared outburst discussed above, no clear
statement can be made for the preceding ejections.

\cite{Abrahametal96} have also correlated the epoch of ejection of components with
the radio light curve at 22\,GHz they claim that the ejection times of 
all components are related to increases in the radio flux. They do
not, however, provide a quantitative assessment of this
relationship. Flux increases are indeed expected to be associated with 
the appearance of new jet components if these are new ejecta that
become optically thin as they move away from the core. A further
possible link has been established by \cite{Krichbaumetal96} between
the ejection of the knots and the high energy activity of 3C~273 as
evidenced by ECRET data.

\sbox{\MaBoiteC}{\cite{Abrahametal96}}
\begin{figure}
\includegraphics[totalheight=15cm,width=12cm]{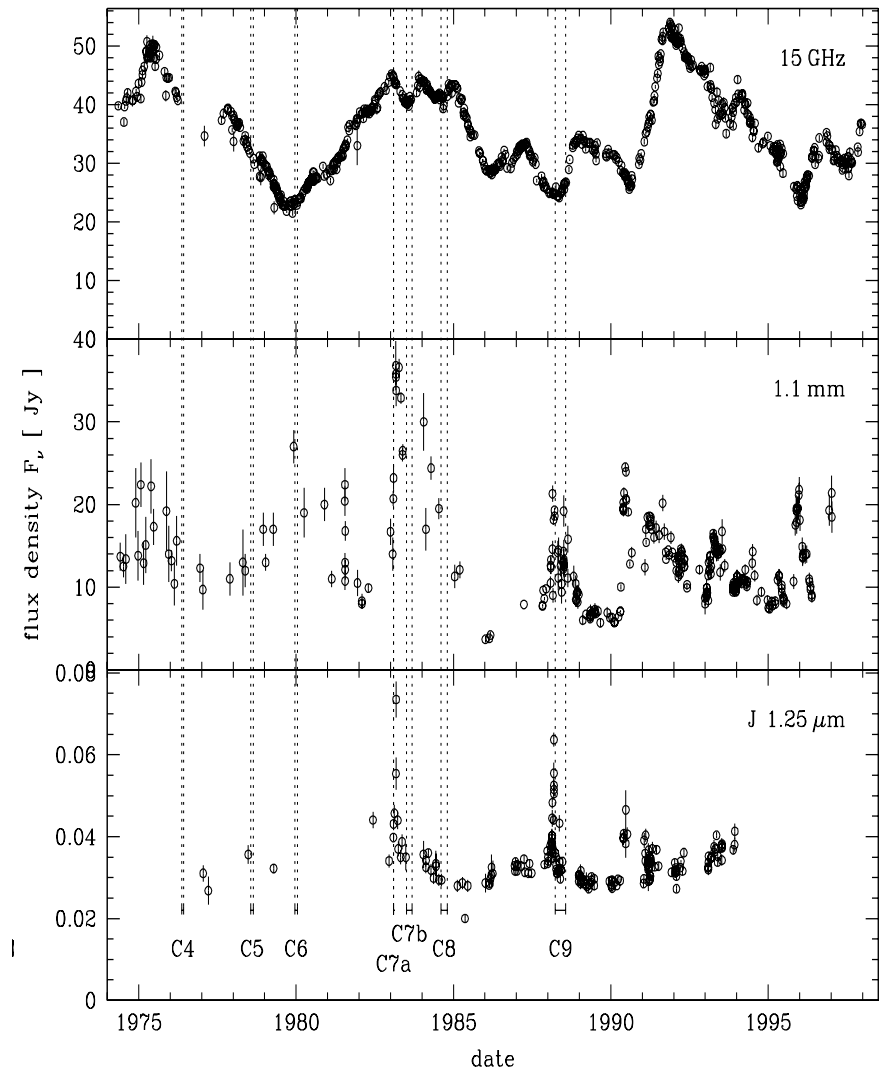}
\caption{\label{fig:mmandinfrared}millimetre and infrared light
  curves and dates of appearance of new VLBI jet components (see the
  text). The components are labeled as in \usebox{\MaBoiteC}. The epochs of
  ejection of the components are from \usebox{\MaBoiteC}. The
  uncertainty in the ejection epochs are shown by a short range.}
\end{figure}

The VLBI observations quoted in \cite{Krichbaumetal90} were made at
43\,GHz. VLBI observations at even higher frequencies (100\,GHz) were
obtained by \cite{Baathetal91}. These data reach a resolution of
50\,micro seconds of arc, illustrating the power of the technique.
Using $H_0=50$\,km/(s\,Mpc) this angular resolution corresponds to a
linear scale of 5\,$10^{17}$\,cm at the distance of 3C~273. This is to
be compared with the gravitational radius of a $10^{10}$ solar masses
black hole, which is 3\,$10^{15}$\,cm. In other words, modern VLBI
observations are capable of resolving structures in the radio data of
3C~273 down to 100 gravitational radii. This effort to obtain maps at
higher frequencies is being pursured (see e.g.~\cite{Krichbaumetal97}).

High angular resolution VLBI data reveal that the angle at which the
jet emerges from the core is significantly different at the hundred
micro arcsecond scale ($-119^\circ$) from that observed at the mili
arcsecond scale ($-130^\circ$) or at longer scales ($-137^{\circ}$)
\cite{Baathetal91} and references therein). \cite{Baathetal91}
interpret this result as being due to either deflection of the jet or
(but this is in a sense equivalent) to changes in the speed of the
jet. This must be put in parallel with the observation of
\cite{Krichbaumetal90} who report that the velocities of the
individual knots are different (from 4$\pm 0.3$ to 8$\pm 0.2$ times
the velocity of light).

Another type of improvement in the knowledge of the jet was brought about by
investigations with a higher dynamical range. Such observations are reported in
\cite{Davisetal91}. Two important results follow from their data. The superluminal motions observed at small distances from the core
extend to at least 240\,pc ($H_0=50$\,km/(s $\cdot$ Mpc)). The
velocity at large distances is only marginally less than closer to the
core. The second result is that no counter
jet is detected. The brightness ratio between a jet and an
intrinsically identical counterjet is given by

$$[(1+\beta \cos \theta)/(1-\beta \cos \theta)]^{2+\alpha}.$$

Using a spectral index of 0.8 \cite{Davisetal91} deduce from the
observed lower limit on this ratio that $\beta \geq 0.95$. This
velocity is close to that obtained from the superluminal expansion
(see below). The data available is therefore still compatible with the
presence of a counter jet of similar properties to the one observed
but unobserved due to its relativistic motion away from us. A further
improvement of the dynamical range by a factor of a few would provide an
important set of data on the intrinsic properties of an eventual
counter jet.

The intrinsic velocity of a relativistic jet can be deduced from the
apparent proper motion in the following way (the original model is due
to \cite{Blandfordetal77}):
\begin{figure}
\hspace{7pt}
\includegraphics[viewport=19bp 0bp 484bp 772bp,scale=0.46,angle=-90]{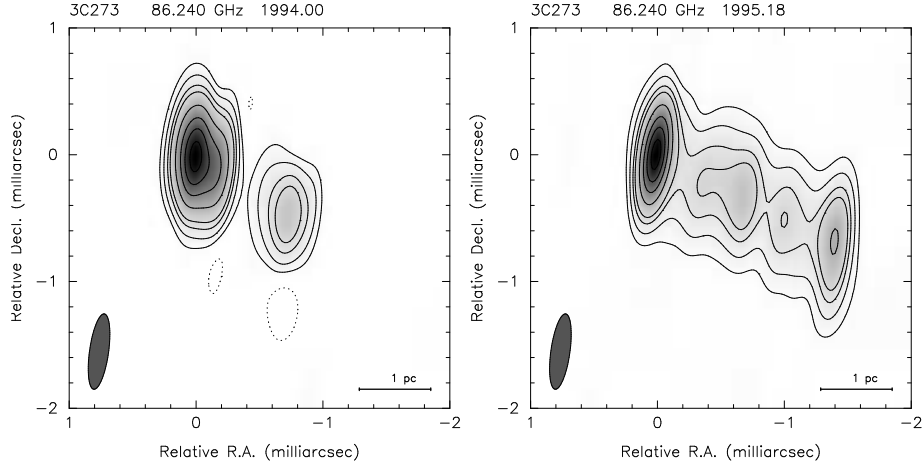}
\caption{\label{fig:Geometry_of_jet}The VLBI Jet of 3C~273 at two
  different epochs in 1994 and 1995 observed at
  86\,GHz. (courtesy T. Krichbaum.)}
\end{figure}
  
Assume that a photon is emitted by a blob of the jet that has traveled
during $\delta t$ at the velocity v. The difference in arrival time of
this photon and one that originated from the base of the jet at the
time of departure of the blob $\Delta t$ is

$$\Delta t = \delta t (1- v/c \cos\theta)\,.$$

The motion of the blob a perpendicular to the line of sight seen by an
observer far away is

$$v_{\perp} = \frac{\delta t v \sin\theta}{\Delta t}\,.$$

It is easily seen from both expressions that superluminal motion can
be observed for $v$ close to c and $\cos\theta$ close to one and that
the angle $\theta$ for which the transverse velocity is
maximum for a given intrinsic velocity is given by

$$\cos\theta_{max} = v/c \,.$$

\section{The Host Galaxy}

If there is one subject for which the brightness of 3C~273 is a
problem rather than a help it is the study of the underlying galaxy.
Subtraction of the quasar light distribution on images has proved a
very delicate issue and has been possible only after the introduction
of CCD cameras. \cite{Kristian73} followed the idea that the quasar
phenomenon is similar in nature to that observed in the nuclei of
Seyfert galaxies but considerably more powerful and studied whether
galaxies can be seen around at least some quasars in direct
photographic plates. He concluded that in some cases galaxies are seen
while for the brighter objects (including 3C~273) this is not the
case, a conclusion expected from the quality of the data.

First (to my knowledge) CCD images of 3C~273 are presented by
\cite{Tysonetal82}. These authors conclude that there is an underlying
galaxy and that it is similar to the giant elliptical galaxy NGC~4889
in the Coma cluster. The integrated V magnitude of the nebulosity is
about 16.05, 3 magnitudes less than the quasar. Based on some previous
data on the field of 3C~273 by \cite{Stockton80} they also concluded
that the 3C~273 host galaxy may be a member of a poor cluster of
galaxies.

Near infrared data may be of interest to study the galaxies around
quasars as the ratio of galaxy to QSO fluxes may be in general larger
than at visual wavelengths (quasars are blue
objects).\cite{veroncat90} show thus that for luminous quasars there
is no correletion between the luminosity of the galaxy and that of the
QSO. \cite{McLeodRieke94} present a sample of high luminosity quasars
including 3C~273. They show that there is in their sample some
correlation between the QSO luminosity and that of the underlying
galaxies, 3C~273 having the most luminous underlying galaxy of their
sample and the smallest fraction of galaxy to (QSO+galaxy) luminosity.
The correlation shows, however, a large dispersion. Using far infrared
luminosities, \cite{McLeodRieke94} also investigate the possible
existence of starburst activity in the galaxies surrounding quasars.
They (preliminarily) conclude that this may be excluded in several of
the galaxies, not in all and in particular not in 3C~273.

HST data have recently been published by \cite{Bahcalletal97} for a
sample of nearby luminous quasars. The main advantage of the
(repaired) HST is, expectedly, the sharpness of its point spread
function that allows a cleaner subtraction of the quasar light from
the image. This still requires, however, a good knowledge of the point
spread function at some distance from the central peak, where the
galaxies contribute most of their light.  It is interesting to see
that the HST data is in contradiction with the popular idea that radio
loud quasars lie in bright elliptical galaxies while the radio quiet
quasars, like Seyfert nuclei, lie in spiral galaxies.  We can conclude
from this that it is not the obvious shape of the galaxy that
determines the characteristics of the quasar.

In the case of 3C~273, the inferences drawn from the HST data confirm
those obtained from the ground. The galaxy may be classified as an E4
galaxy, its luminosity is roughly 3 magnitudes fainter than the quasar
in the visible. This indicates that the galaxy is somewhat brighter
than the most luminous galaxy of a rich cluster. There is no
conspicuous companion or signs of recent violent interaction with
another galaxy. The galaxy major axis is about $30\prime\prime$,
corresponding to some 100\,kpc with the cosmological parameters used
here.

Whereas the study of a single object cannot have universal value, it
is still interesting to note that 3C~273, the brightest nearby quasar,
is not imbedded in a distorted or peculiar galaxy. This shows that
while galaxy interactions may play an important role in bringing
material from the galaxy to its nucleus, some other mechanisms must
also be at work and even have the dominant role in at least some very
bright cases.

\section{Understanding it \label{sect:8}}

The paradigm of QSO physics is that the energy is freed by accretion
of matter in a massive black hole. This paradigm allows some simple
estimates:

The Eddington luminosity for which gravitational attraction
compensates radiation pressure is:

$$L_{Edd} = 1.3 \mathrm{x} 10^{38} \frac{M}{{M}\odot} \mathrm{ergs/s}\,.$$

Using the bolometric luminosity of 3C~273 deduced in
Sect.~\ref{sect:3.5} we thus estimate that provided that the bulk of
the luminosity is emitted isotropically the mass of the central black
hole is $\sim 10^9$ solar masses. The corresponding gravitational
radius is $\sim 3 \cdot 10^{14}$\,cm.

The mass accretion rate can be estimated from the luminosity $L$ by:

$$L = \eta \cdot \dot{M}c^2\,,$$

where $\eta$ is the efficiency of the conversion of rest mass to radiation.

Using the same luminosity as above and an efficiency $\eta$ $\sim$
10\% typical of accretion onto black holes we deduce a mass accretion
rate $\dot{M} \sim 1.3 \cdot 10^{27}$\,g s$^{-1}$ or about 10 solar
masses per year. Not surprisingly, these numbers are close to those
deduced from accretion disc models.

Going beyond these estimates requires understanding of how the energy
liberated by the accretion process is transformed in the radiation we
observe across the electromagnetic spectrum. This understanding is
still widely lacking, we can, however, list some of the elements
that do play a role.

Highly relativistic electrons and magnetic fields must belong to any model
of 3C~273 (and other radio loud AGN) as shown by the presence of
synchrotron radiation. Their energy densities are very inhomogenous and
 partly organized in small packets some of which at least are
accelerated to relativistic bulk velocities along complex paths to
form the observed small scale jet. It is probable that the electrons
are accelerated to highly relativistic energies in shocks and that they thus
aquire the energy that they radiate from the kinetic energy of some
underlying flows.

Another indication of fast flows is the presence of the broad lines
which indicate that the material surrounding the black hole has
velocities of the order of $10^4$\,km/s. It has not been possible to
find in the line variability pattern the signature for a dominantly
ordered velocity field (expansion, accretion or rotation). One,
therefore, concludes that a large fraction of the velocity field
is of a turbulent nature. In these circumstances, the presence of
shocks where streams of matter collide is difficult to avoid. It is
interesting to note that thermalising  Hydrogen gas with bulk
velocities of the order of $10^4$\,km/s will produce a gas of T $\sim
5 \cdot 10^9$\,K. A temperature close to the one needed to Comptonise
the UV photons to X-rays with a slope as
observed~\cite{WalterandC92}. 

A fraction of the UV and higher energy radiation is reprocessed by gas 
to form the broad lines and by dust to give the thermal infrared
radiation.
The organisation of the broad line emitting clouds is unclear, no
cloud confining medium having been found. Whether the optical-UV
emission forming the blue bump is itself due to reprocessed X-ray
emission is also unclear, mostly because of the absence of the
signature of Compton reflection in the X-rays.

There are many different timescales at play. Among those we know there 
is the few days delay between the UV and optical light curves which
imply that the signals ruling the blue bump emission travel at the
speed of light. There is also the presence of much longer timescales
(of several years) in the visible light curves and correlations which
delays of the order of a year or so, between emission
components. These timescales are long compared with light crossing
times or dynamical  times in the vicinity of the black hole. They are,
however, short compared to viscous timescales of standard accretion
discs. \cite{CourvoisierClavel91}. The presence of these timescales
may either indicate that the size of the continuum emitting accretion
is of the order of a parsec (similar to the size of the broad line
region) or that there exist characteristic velocities of the order of few
percent of the speed of light in a region of several gravitational
radii. One may also note that the amplitude of the variations at short 
timescales ( one day $\sim$ 10 gravitational radii over $c$) is small
(few percent; \cite{Paltanietal98}). This may
indicate that the variations on this timescale are not associated
with the regions closest to the black hole, but rather to small
regions in an extended object.

There have been many attempts to understand the geometry of the
emission regions considering one or several emission components. Most
have been based on the presence of accretion discs. Many of the
arguments are revised in \cite{Blandford90}. The addition of a corona
being discussed by \cite{Haardtetal94}.

\cite{CamenzindCourvoisier83} attempted to understand the continuum
emission of 3C~273 in terms of a mildly relativistic wind originating
in the core of the object and shocked at some distance. Most of the
observed emission in this model was the by product of the shocked
material. This model predicted that the variation time scales of the
different components was such that the UV varied faster than the
X-rays which in turn varied faster than the optical emission. The
infrared and gamma ray variability timescales were expected to be the
longest. These predictions were soon disproved by observations which
led to a revision of the geometry \cite{CC89}. In this revised
geometry the wind is channeled in such a way that the shocked material
covers only few percents of the UV source. The shocked material is
heated to temperatures such that the UV photons crossing it are
Comptonised to X-ray energies. The lag between X-ray and UV fluxes may
be understood naturally in this geometry \cite{Paltanietal98}.

\cite{Courvoisieretal96} have considered whether accretion of matter
could be in the form of stars rather than gas. In their model the
gravitational energy is radiated following collisions between stars in
the vicinity of a black hole. First order considerations showed this
to be a possible alternative to understand the variability of AGN and
its dependence on luminosity. This also points to the little studied
question of the interaction between the active nucleus and the
surrounding stellar population.

It is thus clear that although the main elements of the AGN model have 
been in place for more than 30 years, often following pioneering
observations of 3C~273, much remains to be understood. AGN are
considerably more complex than many of us anticipated. This complexity 
together with the extreme properties they show make them fascinating
object to study.


\section*{Acknowledgments}

This work is based on a long term effort by a large set of colleagues
who have participated in the gathering of data and in many discussions
over the years. I owe a particular debt to M. Camenzind and M.-H.
Ulrich for sharing their knowledge with me when this effort began. I
would never have been able to write this review without the benefit
from many interactions over the years and around the world. Several of
my colleagues at the ISDC have given me some very direct help in
preparing this review and in particular the figures. They are S.
Paltani, M. Polletta and M.  T³rler. I thank them and also T.
Krichbaum, R. Walter and L. Woltjer for reading and commenting the
manuscript. A. Aubord and M. Logossou have been of much help in the
typesetting.  \input{biblio}

\end{document}
